%% file: sample-sigconf.tex
\documentclass[sigconf]{acmart}

\usepackage{soul}
\usepackage{xspace}
\usepackage{listings}
\usepackage{xcolor}
\usepackage{booktabs}
\usepackage{cleveref}
\usepackage[ruled,linesnumbered,vlined]{algorithm2e}
\usepackage{algorithmic}
\usepackage{tikz}
\usetikzlibrary{arrows, arrows.meta, patterns}
\usepackage{pifont}
\usepackage{caption}
\usetikzlibrary{positioning}
\usepackage{smartdiagram}
\usesmartdiagramlibrary{additions}
\usepackage{enumitem}
\usepackage{amsthm}
\usepackage{csquotes}
\usepackage{tcolorbox}
\tcbuselibrary{theorems, skins, breakable}
\usepackage[normalem]{ulem}
\usepackage{array}
\usepackage{fancyhdr}
\SetAlgoSkip{-10pt}

\usepackage{titlesec}
\titlespacing*{\section}{0pt}{1.25ex}{1ex}
\titlespacing*{\subsection}{0pt}{1.1ex}{1ex}
\titlespacing*{\subsubsection}{0pt}{1ex}{1ex}

\definecolor{yamlkeycolor}{RGB}{18,123,169}    %
\definecolor{yamlstringcolor}{RGB}{80,143,79}  %
\definecolor{yamlcommentcolor}{RGB}{128,128,128} %
\definecolor{yamlbackground}{RGB}{250,250,250} %
\definecolor{yamlframe}{RGB}{220,220,220}      %

\lstdefinestyle{yamlstyle}{
    basicstyle=\ttfamily\scriptsize,
    backgroundcolor=\color{yamlbackground},
    breakatwhitespace=true,
    breaklines=true,
    captionpos=b,
    commentstyle=\color{yamlcommentcolor},
    deletekeywords={...},
    escapeinside={\%*}{*)},
    extendedchars=true,
    frame=single,
    framesep=2pt,
    keepspaces=true,
    keywordstyle=\color{yamlkeycolor}\bfseries,
    morekeywords={type, output, schema, prompt},
    numbers=left,
    numbersep=5pt,
    numberstyle=\tiny\color{yamlcommentcolor},
    rulecolor=\color{yamlframe},
    showspaces=false,
    showstringspaces=false,
    showtabs=false,
    stepnumber=1,
    stringstyle=\color{yamlstringcolor},
    tabsize=1,
    xleftmargin=10pt,
    xrightmargin=2pt,
    framexleftmargin=10pt,
    framexrightmargin=2pt,
    framexbottommargin=2pt,
    framextopmargin=2pt,
    columns=flexible,
    linewidth=\columnwidth,
    basewidth=0.3em,
    lineskip=0.5pt,
    literate=
     *{:}{{\textcolor{yamlkeycolor}{:}}}{1}
     {|}{\textcolor{yamlkeycolor}{|}}{1}
     {\ \ }{{\ \ }}{1},
    emph={type,output,schema,prompt,validate,num\_retries\_on\_validate\_failure,operations,pipelines,input,model,join\_key,comparison\_model,comparison\_prompt,reduce\_key,resolution\_prompt,resolution\_model,output\_keys,prompts,content\_key,peripheral\_chunks,split\_key,chunk\_size,main\_chunk\_start,main\_chunk\_end,gleaning,unnest\_key,expand\_fields,datasets},
    emphstyle={\color{yamlkeycolor}\bfseries}
}

\lstdefinestyle{plaintextstyle}{
  breaklines=true,
  breakatwhitespace=false,
  basicstyle=\footnotesize\ttfamily,
  columns=flexible,
  keepspaces=true,
  showstringspaces=false,
  frame=single,
  framesep=3pt,
  numbers=none,
  xleftmargin=0pt,  %
  xrightmargin=3pt,
  resetmargins=true, %
  aboveskip=10pt,    %
  belowskip=10pt,    %
  gobble=0           %
}

\lstdefinelanguage{text}{
  identifierstyle=,
  keywordstyle=,
  commentstyle=,
  stringstyle=,
  morekeywords={}
}

\lstnewenvironment{yaml}[1][]
  {\lstset{style=yamlstyle, #1}}
  {}

\AtBeginDocument{%
  }

\renewcommand\footnotetextcopyrightpermission[1]{} %

\setcopyright{none}
\acmYear{2024}
\acmDOI{XXXXXXX.XXXXXXX}

\input{macros}

\begin{document}

\setcounter{page}{0}

\title{\docetl: Agentic Query Rewriting and Evaluation \\ for Complex Document Processing}

\author{Shreya Shankar$^1$, Tristan Chambers$^2$, Tarak Shah$^2$, Aditya G. Parameswaran$^1$, Eugene Wu$^3$}
\affiliation{%
$^1$UC Berkeley EECS, $^2$BIDS Police Records Access Project, $^3$Columbia University \\
\{\url{shreyashankar, tristan.chambers, tarak_shah, adityagp}\} \url{@ berkeley.edu}, \url{ewu @ cs.columbia.edu}\country{}}

\renewcommand{\shortauthors}{Shankar et al.}

\input{sections/abstract}

\maketitle
\pagestyle{plain} %

\input{sections/intro-agpSep10}

\input{sections/dsl}
\input{sections/rewrite}
\input{sections/optimizer}

\input{sections/new_eval}

\input{sections/discussion}

\input{sections/related}
\input{sections/conclusion}

\bibliographystyle{ACM-Reference-Format}
\bibliography{sample-base}

\clearpage
\appendix

\input{sections/appendix}

\end{document}

%% file: macros.tex
\newcommand{\techreport}[1]{#1}
\newcommand{\papertext}[1]{}

\newcommand{\topic}[1]{\vspace{-3.5pt}\smallskip \smallskip \noindent{\bf #1.}}
\newcommand{\docetl}{\textsc{DocETL}\xspace}
\newcommand{\ttt}[1]{{\small \texttt{#1}}\xspace}

\newcommand{\revision}[1]{{#1}}

\newif\ifrebuttal
\rebuttalfalse %

\ifrebuttal
\newcommand{\revision}[1]{\textcolor{blue}{#1}}
\fi

%% file: sections/abstract.tex
\begin{abstract}
Analyzing unstructured data 
has been a persistent challenge in data processing. 
Large Language Models (LLMs) have shown promise in this regard, 
leading to recent proposals for declarative
frameworks
for LLM-powered processing of unstructured data. 
However, these frameworks 
focus on reducing cost when executing user-specified
operations using LLMs, rather than improving accuracy, 
executing most operations as-is (in a single LLM call).
This is problematic 
for complex tasks and data, where LLM outputs for user-defined operations 
are often inaccurate, even with optimized prompts. 
For example, an LLM may struggle to identify 
{\em all} instances of specific clauses, like force majeure or indemnification, in lengthy legal documents, requiring decomposition of the data, the task, or both. 

We present \docetl, a system that optimizes complex document processing pipelines, 
while accounting for LLM shortcomings. \docetl offers a declarative interface for users to define such pipelines and uses an agent-based approach to automatically optimize them,
leveraging novel agent-based rewrites (that we call {\em rewrite directives}), as well as an optimization and evaluation framework. 
We introduce {\em (i)} logical rewriting of pipelines, tailored for LLM-based tasks, {\em (ii)} an agent-guided plan evaluation mechanism that synthesizes and orchestrates task-specific validation prompts, and {\em (iii)} an optimization algorithm that efficiently finds promising plans, considering the latencies of agent-based plan generation and evaluation. Our evaluation on four different unstructured document analysis tasks demonstrates that \docetl finds plans with outputs
that are $21$ to $80\%$ more accurate than well-engineered baselines. \docetl is open-source at \ttt{docetl.org}, and as of \revision{March 2025}, has amassed over \revision{1.7k} GitHub Stars, with users spanning a variety of domains.
\end{abstract}

%% file: sections/intro-agpSep10.tex
\section{Introduction}
\label{sec:intro}

\begin{figure*}
\papertext{\vspace{-35pt}}
\techreport{\vspace{-10pt}}
    \centering
    \includegraphics[width=0.8\linewidth]{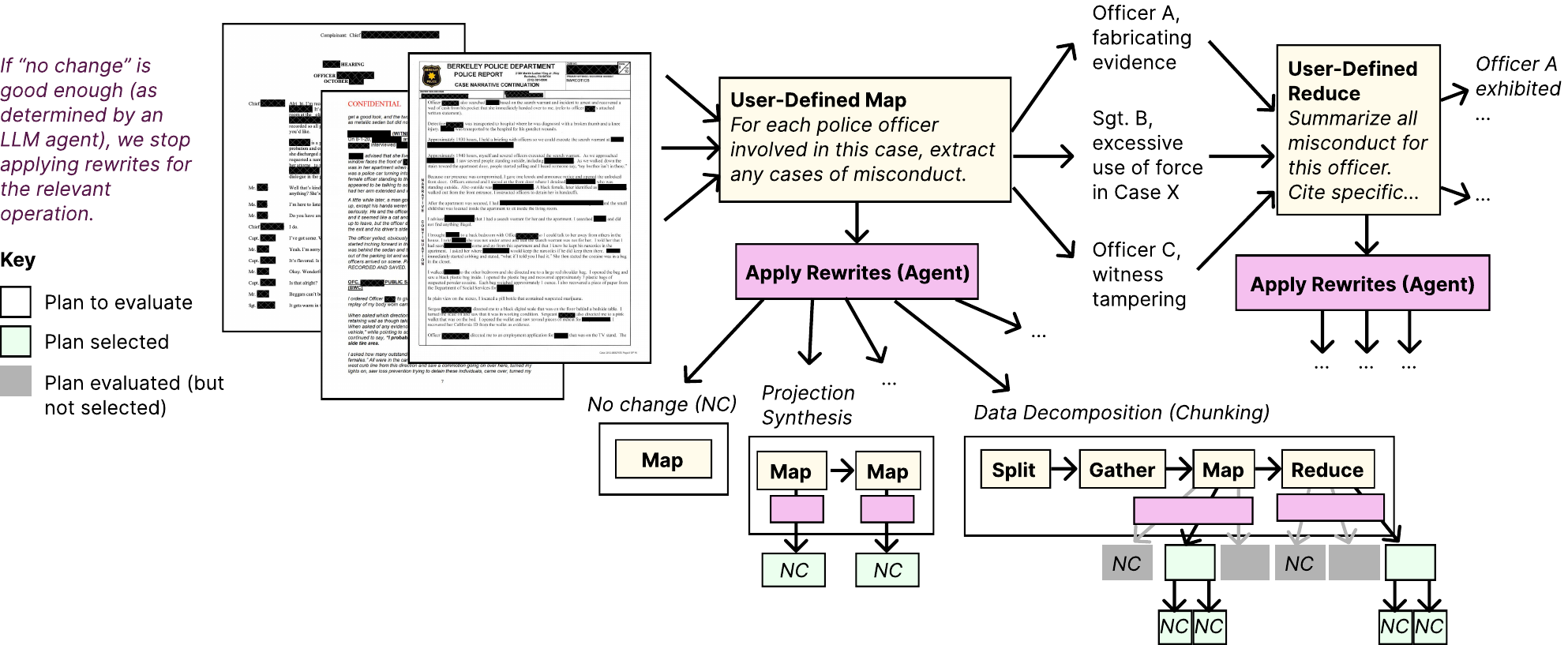}
    \vspace{-5pt}
     \caption{Optimization for a pipeline designed to accomplish the task in \Cref{ex:journalist-task}. The diagram illustrates the system mid-optimization of the initial map operation. \docetl employs LLMs to synthesize new plans using novel rewrite directives. The process begins with an LLM verifier determining if an operation is sufficiently optimized. If not, rewriting continues. Notably, when a new operation is synthesized as part of a rewrite, it undergoes immediate opportunistic optimization, as shown by the nested ``Apply Rewrites (Agent)'' rectangles.}
    \label{fig:optimizer-main}
    \papertext{\vspace{-15pt}}
\end{figure*}

Large Language Models (LLMs) have taken the world of 
data management by storm, with applications 
ranging from data integration,
to tuning, to query optimization, to data cleaning~\cite{llmsdisruptdatamanagement}.
There has also been an interest, all in the last few months, in declarative approaches
to process unstructured data using LLMs~\cite{lin2024towards, patel2024lotus, liu2024declarative, anderson2024designllmpoweredunstructuredanalytics}. 
These systems,
instrumented as extensions to the
relational model
for processing textual columns,
typically assume the text snippets per row
are {\em small and easy to process}.
They therefore focus on reducing cost, while 
keeping accuracy almost the
same. 
However, for many real-world tasks, that
we refer to as {\bf \em  complex document processing} tasks, 
accuracy can be a significant bottleneck, limiting practical utility.
Here, complexity can stem from the 
documents or the nature of the processing task, or 
both.
Consider this scenario from our collaborators
on the Police Records Access Project\footnote{\url{https://bids.berkeley.edu/california-police-records-access-project}}:

\begin{example}[Police Misconduct Identification]
\label{ex:journalist-task}
Journalists at Berkeley's Investigative Reporting Program want to analyze a large corpus of heterogeneous police records, obtained through records requests, to uncover patterns of misconduct and procedural violations. Records include police reports, court transcripts, internal affairs and medical examiner reports, and other case files, often spanning hundreds of pages each. Analysis involves extracting key information from long documents, aggregating information across documents to identify behavioral patterns for each officer, and generating summaries highlighting concerning trends.
\end{example}

\noindent \Cref{ex:journalist-task} is representative of  
complex document processing tasks 
across domains including law, medicine, 
and social science. 
Consider a simpler version of this task,
where we just want a summary 
of the role of each officer 
mentioned in each complex police record
document, each with hundreds of pages. 
This task can be expressed as a 
single-step {\em map}
operation applied to the OCR output per document, in one LLM call,
with a user-provided 
prompt defining terms like ``misconduct.''  
All existing systems~\cite{lin2024towards, patel2024lotus, liu2024declarative, anderson2024designllmpoweredunstructuredanalytics} 
would simply execute the map operation, as is, with one LLM call per document. 
That is, they {\bf \em assume user-defined operations
will yield sufficiently accurate results when executed by the LLM},
and focus primarily on reducing cost.
However, 
this map operation may provide poor accuracy for multiple reasons.
First, the document in question may exceed the LLM's context limit. Even if it fits, outputs may omit certain
instances of misconduct, or
include spurious information.
Recent work has shown that 
LLM performance degrades considerably as length increases~\cite{levy2024same},
because they can be distracted~\cite{shi2023large} or
selectively pay attention to certain portions~\cite{liu2024lost}, failing to gain a holistic understanding~\cite{bai2023longbench,tang2023found,zhao2024longagent,jiang2023longllmlingua}.
Simultaneous
theoretical work has shown that this degradation is due
to limits in the transformer architecture~\cite{peng2024limitations, sui2024confabulation, kalai2024calibrated}.
While one could apply prompt compilation~\cite{khattab2024dspy,wen2024hard}
to identify a better prompt,
this relies on examples, which are
either not present or are too long to include (e.g., 
an example document with hundreds of pages)---but
irrespective do not fix the underlying challenges with LLMs performing
a complex task on complex documents.

Our key insight is that 
the quality of LLM outputs is often not adequate
for complex data processing---we cannot simply treat the existing user-provided operators as fixed.  Instead, 
{\bf \em we need to consider novel rewrites that decompose
complex but error-prone operation(s) 
into a sequence of simpler and more accurate operations.}
For our map example, 
a different sequence of operations may increase accuracy.
One such example is {\em map $\rightarrow$ map}, where the first map
is tasked with removing all portions of 
each input document that do not pertain
to misconduct (e.g., medical reports),
while the second map is the single-step map above. 
Or we could replace the first map
with one that summarizes each sequence of $k$ paragraphs into one,
keeping the second map as is. 
Yet another option is to replace the single-step {\em map} with
what we call {\em split $\rightarrow$ gather $\rightarrow$ map $\rightarrow$ reduce}---a pattern that first {\em splits} the document into contiguous chunks;
then, for each chunk, {\em gathers} $k$ neighboring chunks before/after as context or background 
to be included into a prompt, generates per-officer summaries using its $2k$ neighbors as background context ({\em map});
and finally, performs a global summarization across all chunks ({\em reduce}).

However, {\bf \em we cannot expect a user to rewrite their pipeline into multiple alternatives and determine the one with the best performance}. The previous paragraph introduced three out of a multitude of potential rewrites, each of which could be recursively applied to operators in a pipeline, presenting a seemingly infinite set of options.
For example, for the {\em map $\rightarrow$ map} pipeline, 
there are many alternatives for what the first map could do,
and many different associated prompts. 
Even if we decide to use the first map to summarize $k$ chunks at a time,
determining the right value for $k$ is challenging.
Likewise for {\em split $\rightarrow$ gather $\rightarrow$ map $\rightarrow$ reduce}. 
Moreover, we're just focusing on the first step
of the overall goal in \Cref{ex:journalist-task},
which is to summarize misconduct across all documents. 
So, we may need to apply a {\em reduce}
operation across documents 
to group and summarize misconduct extractions by officer. 
However, the same officer may be extracted as 
``Officer Smith'' in one document and ``J. Smith'' in another, resulting in separate, incomplete summaries for what should be a single officer~\cite{parameswaran2023revisiting}.
It's not entirely clear how one would implement
this form of entity resolution, and no current systems support it. 
In fact, additional context from the original document(s) may be necessary
to determine if the two officers with the same name are identical.
Finally, LLMs might struggle to recognize that multiple documents are from the same case, leading to overrepresentation of incidents in the misconduct summaries~\cite{Schaik2024AFG}. 
{\bf \em Overall, even an LLM expert would need extensive experimentation to design an accurate pipeline, given the dependency on the data, task, and LLM capabilities.} 
This complexity underscores the need for a system 
that can automatically explore and evaluate different 
task decomposition strategies 
to find the most effective pipeline 
for a given task and dataset.

We present \docetl, our first attempt at developing 
a {\bf \em declarative system optimized for accurate complex document processing}.
\docetl provides a declarative YAML-based interface for users
to author pipelines with LLM-specific operators, 
including two new ones: {\em resolve} for entity resolution, and {\em gather} to maintain context when processing document chunks. 
Users can specify their pipeline
at a high level with \docetl decomposing, rewriting, and optimizing
the pipeline. 
\docetl introduces an {\em agent-based framework}
to {\em rewrite} user-specified pipelines into alternative ones, 
as shown in \Cref{fig:optimizer-main}.
Rather than simply relying on agents as-is, which can be error-prone, we
{\em guide} them to rewrite query plans using novel {\bf \em rewrite directives} that we identify. 
We call these directives instead of rules because they are abstract guidelines interpreted by LLMs based on task and data characteristics, with infinitely many concrete instantiations.
We further leverage an agentic framework to 
{\em evaluate} the resulting pipelines.
Since evaluation can be expensive,
we develop an optimization approach
inspired by Cascades~\cite{Graefe1995TheCF},
where we use a top-down rule-based strategy
to generate and evaluate a space of equivalent plans,
opting to opportunistically decompose (or rewrite)
complex or error-prone operations
into simpler ones.

\docetl is open-source and available on GitHub\footnote{\ttt{https://github.com/ucbepic/docetl}}.
As of \revision{March 2025}, it has already amassed \revision{\bf 1.7k+ GitHub stars},
and has been used for pipelines ranging from domain-specific analysis (e.g., legal, climate science) to enterprise and personal productivity (e.g., analyzing customer support tickets, emails);
over \revision{{\bf 400 users}} have joined the corresponding Discord server. 

Overall, finding optimal complex data processing pipelines is impossible given the infinite search space, non-determinism of LLMs, fuzziness of text, and ambiguity in task-specific success criteria. However, even in these difficult settings, \docetl is able to produce pipelines that are sufficiently accurate for practical needs, as is evidenced by our adoption across domains.  \docetl is able to do so by leveraging the power of LLM agents in constrained ways, in conjunction with a powerful, but compact set of rewrite directives, decomposition into processing units that can be validated, as well as an opportunistic top-down exploration of the search space.

We make the following contributions in this paper:

\begin{enumerate}[nosep, leftmargin=*, wide=0pt]

\item {\bf Novel Rewrite Directives and Agent-Driven Rewriting:} We identify 13 new rewrite directives designed for LLM-based operators, 
addressing challenges unique to complex document processing.
Unlike traditional rewrite rules, 
LLM agents are used to implement these directives. 
When a rule applies to a portion of a pipeline, 
agents synthesize appropriate prompts and parameters for new operations. 
For example, when decomposing a ``summarize instances of misconduct'' operation into multiple ones, an agent might create two steps: first, ``list instances of misconduct given specific types (e.g., excessive force),'' 
followed by ``summarize each listed instance,'' 
crafting suitable prompts for each new operation.

\item {\bf Agent-Driven Plan Assessment:}  We also use LLM agents to synthesize task-specific validation prompts for each operation, which are used to assess output quality. 
For instance, to verify a misconduct summary, 
an agent might create a prompt, 
``Does this summary include {\em all} instances of misconduct from the document?'' Or, ``Do all mentioned instances {\em actually exist} in the document?'' 
The agents then execute plans on sample data 
and evaluate outputs using these custom prompts. 
This entire process happens without the user having to provide
or manually validate examples.

\item {\bf Opportunistic Sub-plan Optimization:} 
Unlike traditional query optimizers that generate 
and evaluate a broad range of possible plans~\cite{chaudhuri1998overview}, we leverage an opportunistic top-down search strategy as shown in \Cref{fig:optimizer-main}: 
when we use a rewrite directive to decompose operators
into new ones, we immediately optimize
each new operator. 
We first check if each such 
operator is sufficiently accurate,
based on the validation as described previously. 
If sufficiently accurate, we no longer optimize that operator,
focusing instead on rewriting others. 
Thus, we opportunistically decompose (or apply rewrite directives to) operators
that are not sufficiently accurate,
Such an approach is necessary because enumerating and 
evaluating all theoretically-possible plans would be prohibitively 
time-consuming due to the inherent latencies in LLM operations.
\end{enumerate}

We describe \docetl's programming model and operators in Section~\ref{sec:system-overview};
our new LLM-centric rewrite directives in Section~\ref{sec:rewrite},
the agentic optimizer that applies them, and evaluates the
resulting plans, as well
as the overall framework for optimization in Section~\ref{sec:optimizer}.
We present our initial evaluation in Section~\ref{sec:evaluation}, where we demonstrate that across four unstructured document analysis tasks, \docetl finds plans that are {\bf 21 to 80\% more accurate} than baselines. We\techreport{ then reflect on next steps in Section~\ref{sec:discussion}, and} discuss related work in Section~\ref{sec:related}.

%% file: sections/dsl.tex
\section{\docetl DSL and Operators}
\label{sec:system-overview}

This section presents \docetl's programming model and operators.

\subsection{Programming Model}
\label{sec:overview-programming-model}

\docetl processes collections of documents. A {\em document} comprises a set (or dictionary) of key (or equivalently, attribute)-value pairs, represented as a JSON object. For example, a police record could be a set of key-value pairs, where one key corresponds to the OCR output of the PDF,
while other keys could capture metadata such as agency, file name, or creation date.
A collection of documents or {\em dataset}, is a JSON array. This data representation lets us handle various data types and degrees of structure and easily reference data within operation prompts.  Documents can be nested, e.g., a police record may contain an array of \ttt{related\_documents} that each contain witness statements or evidence logs that are further nested.

\topic{\docetl DSL} \docetl employs YAML as its domain-specific language (DSL) to define data processing pipelines, for several reasons. 
First, YAML is flexible in accommodating complex multi-line 
prompts and examples, as well as output schemas and validation mechanisms, 
while intermixing formatting with arguments in Jinja~\cite{jinja}.
Second, YAML is human-readable and doesn't require extensive coding expertise.
Third, it is commonly used in industry for describing data pipelines (Apache Airflow, dbt, Prefect) 
and services (Kubernetes, Docker, Circle/Gitlab CI/CD). 
Finally, YAML serves as a simple intermediate format for 
representing the \docetl-optimized pipelines 
for human inspection, as well as for 
our no-code interface\techreport{, 
where users will provide data and natural language descriptions,
with \docetl generating optimized pipelines}.
That said, our optimization techniques are not dependent on YAML
and are also applicable to other frameworks.

\topic{\docetl Pipelines} A \docetl \ {\em pipeline}, expressed in YAML, describes a sequence of {\em operations}. Each operation specifies its operator type, input source, prompt template, and output schema. The input source can be either the original dataset or the output of a previous operator.  We refer to this input using pre-defined variables \texttt{input} or \texttt{inputs} depending on whether the input cardinality is one or many. \techreport{A global default model can be specified, and individual operators can override this setting.} The pipeline begins with dataset definitions, which serves as the initial input. As operators process data, they generate output obeying their schemas, which subsequent operators can then use. This structure allows for flexible and modular pipeline composition.   \docetl supports a default model for the entire pipeline, with the option for per-operation model specifications.

\topic{Fault Tolerance} When executing an LLM-powered operator for many input documents in a pipeline, some operations may {\em occasionally} fail to adhere to the given prompt. 
While prior work assumes 
reliability in LLM outputs~\cite{anderson2024designllmpoweredunstructuredanalytics, liu2024declarative, patel2024lotus}, \docetl explicitly addresses this variability: for each operator, users can specify validations as Python statements that evaluate to true or false, referencing document and output attributes. If any validation fails, the operation retries, using context from the failure to improve the likelihood of success in subsequent attempts.

\begin{table*}[t]
\vspace{-30pt}
\scriptsize
\caption{DocETL's operator suite, divided into operators that leverage LLMs for semantic processing and auxiliary operators ($^*$) that handle data manipulation. For each operator, we show the required user configuration and a high-level description of its functionality.}
\vspace{-10pt}
\begin{tabular}{llp{0.7\textwidth}}
\toprule
Operator & User Configuration & Description \\
\midrule
Map & Prompt, output schema & Uses an LLM to execute a transformation per document, adding resulting new keys to the schema (and optionally omitting existing ones). \\
Parallel Map & Multiple prompts, output schemas & Uses an LLM to execute multiple independent transformations on each document in parallel, adding the new keys to the schema. \\
Reduce & Group-by keys, prompt, output schema & Uses an LLM to aggregate groups of documents sharing the same key values into one new document per distinct value. \\
Filter & Prompt returning boolean & Uses an LLM to evaluate a condition per document, retaining only those where the condition is true. \\
Resolve & Comparison prompt, resolution prompt & Uses an LLM to identify values for a given key(s) that fuzzily match across documents and generate canonical versions per group of values, replacing them in-place in the documents. \\
Equijoin & Comparison prompt & Uses an LLM to determine if pairs of documents from two datasets should be joined based on fuzzy/semantic matching of the corresponding keys. \\
Unnest* & Array/dict field to unnest & Flattens nested data structures by either creating separate documents from array elements or merging nested dictionary fields into parent documents. \\
Split* & Split key, chunk size & Divides documents into smaller chunks based on token count or other criteria, creating as many new docs as there are chunks. \\
Gather* & Context window configuration & Augments each chunk with context from surrounding chunks based on specified configuration (e.g., previous and next chunk counts), keeping the set of documents the same. \\
\bottomrule
\end{tabular}
\vspace{-10pt}
\label{docetl-operators}
\end{table*}

\subsection{LLM-Powered Operators}
Here, we describe the LLM-powered operators in \docetl\techreport{ and any specific implementation details for executing them with LLMs}. \Cref{docetl-operators} summarizes our operators; detailed syntax can be found in our documentation\footnote{\url{https://www.docetl.org/}}\papertext{, and more thorough descriptions can be found in our tech report~\cite{shankar2024docetl}}.
Most operators are LLM-versions of classic data processing operators, however, we introduce a new {\it resolve} operator, 
used to canonicalize variations in specific attribute values.  
In the following, for succinctness of description, 
we often conflate a {\em document}---a JSON object comprising key-value pairs and the basic unit of processing in a dataset with its {\em textual content}, typically a value for a specific key within
the JSON object.

\subsubsection{Map} \label{sec:dsl-map} The map operator applies an LLM-powered projection, also known as a {\em semantic projection}, to each document in the dataset. Let's consider an example of a map operation:
\begin{yaml}
- name: extract_officer_misconduct
  type: map
  output:
    schema:
      misconduct: "list[{officer_name: str, misconduct_instance: str}]"
  prompt: |
    Analyze the following police record:
    {{ input.document }}
    Extract any instances of officer misconduct or procedural violations. For each instance, provide the name of the officer involved and a brief description of the misconduct or violation.
\end{yaml}
This operation processes each document independently, using the specified prompt. The output schema is a list of key-value pairs (of officer names and misconduct instances). This flexible, semi-structured output format allows for varying numbers of misconduct instances per document. 
\docetl supports prompts using Jinja2 templates, where ``\ttt{\{\{ input.document \}\}}'' allows for insertion of the current document's content. This functionality permits complex prompts with conditional logic (as we will see later). 
When applied, the map operation adds the new attributes specified in the output schema to the existing document. Users can override this behavior and return a subset of attributes by specifying a \ttt{drop\_keys} list.

\docetl also supports {\em parallel} maps, where multiple independent transformations can be applied in parallel to each document. 
For example, one may extract misconduct while another summarizes relevant policies.   Each operation enriches input documents with new attributes and can run in parallel rather than serially.
\techreport{While users could 
technically use a map to specify a parallel map, in many cases, they already have prompt templates corresponding to two or more independent tasks on the same dataset, and this allows them to not have to coalesce their prompts together.}

\subsubsection{Reduce} The reduce operator aggregates information across {\em multiple} documents based on a set of user-specified keys, ultimately producing {\em one} output document per unique combination of attribute values. \techreport{This operation is particularly useful for consolidating information spread across multiple related documents.} For instance, for reducing police reports, the key set might include \ttt{officer\_name} and \ttt{incident\_date}, allowing for the grouping of all reports involving a specific officer on a particular date. Users can define prompt templates that access the grouped documents via \ttt{\{\{ inputs \}\}} (a list of documents sharing the same key values) and the specific key values for the current group via \ttt{\{\{ reduce\_key \}\}}. By default, reduce operations are assumed to be associative, meaning that the order in which documents are processed does not affect the result. However, if the order is significant, users can specify \ttt{associative: False} in the operation definition.

A challenge arises when any given group of documents is too large for the LLM to correctly process. One could use folding or hierarchical merging to process the data in manageable batches~\cite{condie2010mapreduce, gupta1993maintaining}. In folding, each input is serially processed, with an update to an accumulator (or aggregate), while hierarchical merging recursively aggregates inputs in a tree-like structure. \docetl currently implements a {\em batched folding} approach that starts with an empty accumulator and sequentially folds in batches of more than one document at a time. We chose folding because it permits non-associative reduce operations and maintains the original order of inputs. For example, when summarizing a textbook chapter, \docetl may chunk the text into sections\techreport{(where a chunk is a portion of text that an LLM can reliably process)}, summarize each one, and then employ reduce to summarize the section summaries---a process that requires preserving the original reading order. \docetl automatically determines an optimal fold batch size when building the pipeline.

\begin{figure}
\centering
\vspace{-10pt}
\caption{Reduce's iterative folding over 3 batches of documents.   Each batch takes several documents and the current scratchpad as input (left), and updates the mention counts in the scratchpad and accumulated output of entities mentioned multiple times (right).}
\includegraphics[width=0.8\linewidth]{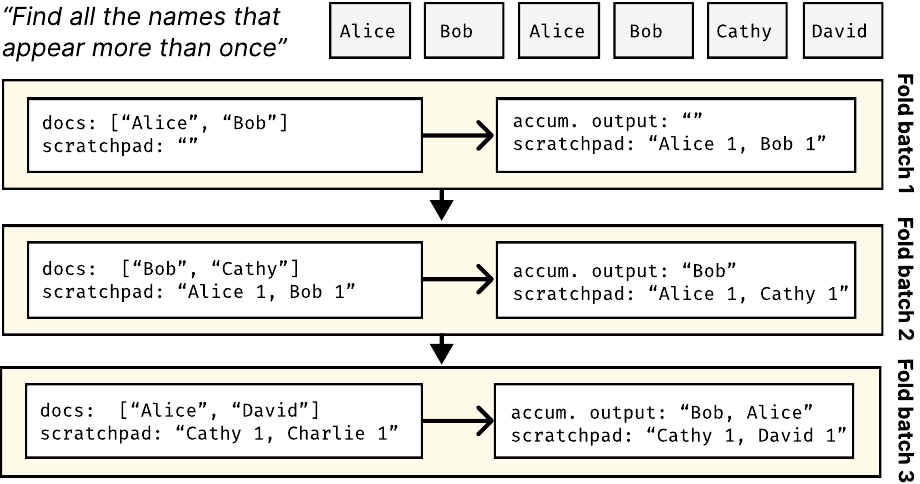}
\label{fig:reduce-folding}
\vspace{-10pt}
\end{figure}

To implement folding, users can provide (or \docetl can generate) a separate {\tt fold\_prompt}, which references the accumulated output and a batch of new inputs to fold into that output. We enhance the system prompt to allow the LLM to write extra notes to a scratchpad~\cite{nye2021show}---a technique that has been shown to improve accuracy by allowing it to maintain state. During each LLM call, we provide the current scratchpad along with the accumulated output and new inputs. The LLM returns both the updated accumulated output and scratchpad, which are passed to the next fold operation. \Cref{fig:reduce-folding} depicts folding for a task to identify names of people mentioned more than once across documents. The scratchpad tracks all mentions of names. As each batch is processed, the LLM updates the scratchpad with new mentions and adds to the accumulated output any person now mentioned more than once.

\subsubsection{Resolve} 

This operator canonicalizes one or more keys across documents that represent slight variations of the same entity. \techreport{for subsequent grouping and aggregation.} Here, resolve reconciles small variations in officer names extracted as part of the map described in~\Cref{sec:dsl-map}:

\begin{yaml}
- name: resolve_officer_names
  type: resolve
  comparison_prompt: |
    Compare the following two officers from police records. Officer 1: {{ input1.officer_name }} mentioned in: {{ input1.record_txt }} and Officer 2: {{ input2.officer_name }} mentioned in: {{ input2.record_txt }} Are these names likely referring to the same officer?
  resolution_prompt: |
    The following names correspond to the same officer:
    {%
      Name: {{ entry.officer_name }}
    {%
    Provide an officer name (first and last) that best represents all the matched entries.
  output:
    schema:
      officer_name: string
\end{yaml}

The user simply specifies how to detect variations, and how to canonicalize them. For instance, ``\ttt{comparison\_prompt}'' checks
whether two officer names are the same, while  ``\ttt{resolution\_prompt}'' chooses a canonical officer name from a list. \docetl
then uses these prompts to compare and resolve 
the officer names. After this operation, the number of documents stays the same. The output schema specifies attributes to replace or add (if new) to each document. Resolve often follows \ttt{unnest} (\Cref{sec:dsl-unnest}), which flattens nested data structures. For example, in our police misconduct pipeline, after unnesting, each document would have distinct \ttt{officer\_name} and \ttt{misconduct\_instance} keys, allowing for name resolution across all mentions in the dataset. Note that users don't need to explicitly define the resolve operation in their pipeline; \docetl will automatically synthesize them if needed 
to ensure consistent entity references across the dataset.
We will discuss how \docetl 
assesses the benefit of such rewrites
in Section~\ref{sec:optimizer-overview}.

\subsubsection{Other Operators}
While expressible using map and reduce,
the following operators are added for convenience. We plan to add
other operators (e.g., sort) in the future. \textbf{Filter} retains documents based on a condition specified in an LLM prompt, which uses a Jinja2 template referencing one or more document keys. \textbf{Equijoin} joins two datasets by comparing documents in pairs, using a \ttt{comparison\_prompt} designed to elicit a binary answer from the LLM, referencing the documents as \ttt{left} and \ttt{right}. The equijoin operation doesn't require an output schema, as the left and right documents are merged to produce the results.

\subsection{Auxiliary Operators}

We present three essential operators that are {\em not} powered by LLMs, used as auxiliary steps to express complex tasks.

\begin{figure*}
\vspace{-15pt}
\centering
\includegraphics[width=0.9\linewidth]{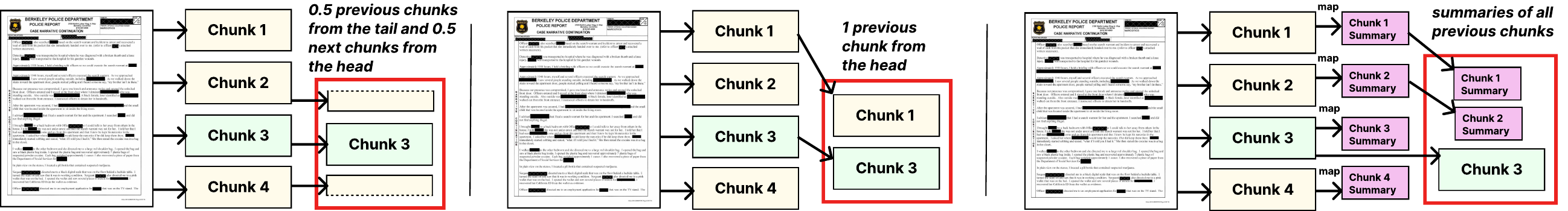}
\vspace{-10pt}
\caption{Split-Gather Pipeline: Illustration of processing a single long document. The split operation divides a long document into manageable chunks. The gather operation then augments each chunk with relevant context from peripheral chunks. The image demonstrates three different ways of rendering chunk 3 (i.e., three different gather configurations): {\em (i)} including fractional parts of surrounding chunks, {\em (ii)} including the full content of the first chunk, and {\em (iii)} including summaries of all previous chunks.}
\label{fig:split-gather-pipeline}
\vspace{-10pt}
\end{figure*}

\subsubsection{Unnest} \label{sec:dsl-unnest} The unnest operator expands an array or dictionary into individual elements. For example, if a map extracts multiple officer names from police interrogation transcripts, each document may contain an array of names. To analyze officers individually across multiple interrogations, unnest creates a separate document for each officer name, effectively flattening the data. This operation can also elevate attributes from nested dictionaries, making them directly accessible for downstream processing.

\subsubsection{Split} \label{sec:overview-utility-split}
The split operator divides long text into smaller chunks. It requires a split key (the text attribute), a split method (token or delimiter), and method-specific parameters (e.g., delimiter or chunk size). \papertext{It generates unique identifiers and sequential numbers for each chunk to enable reassembly later in the pipeline. Resulting documents inherit the other attributes from the original documents.} %
An example is as follows: 

\begin{yaml}
- name: document_splitter
  type: split
  split_key: document_text
  method: token_count
  method_kwargs:
    num_tokens: 1000
\end{yaml}
The above operation splits the \ttt{document\_text} attribute into chunks of 1000 tokens each. The split operation produces several output attributes per chunk:\begin{enumerate}
\item The \ttt{<split\_key>\_chunk} attribute contains the chunk content. Here, the chunk content is stored in \ttt{document\_text\_chunk}.
\item The \ttt{<operation\_name>\_id} attribute contains a unique identifier assigned to each original document (before splitting). In this case, it would be \ttt{doc\_splitter\_id}. All chunks from the same original document share the same ID.
\item The \ttt{<operation\_name>\_chunk\_num} attribute contains the sequential number of each chunk within its original document. Here, it would be \ttt{doc\_splitter\_chunk\_num}.
\end{enumerate} These additional attributes, particularly the document ID and chunk number, are used in downstream gather operations, to reassemble or process the chunks in context. New documents (the result of the split operation) inherit the other attributes from the original documents.

\subsubsection{Gather} \label{sec:dsl-gather} The gather operation complements the split operation by augmenting individual chunks with peripheral information necessary for understanding the chunk's content. Conceptually, gather is similar to windowing in SQL, as both allow ordered access to data beyond the current row or chunk, but gather is specifically designed for LLM-based processing. 
For example, in a transcript split into chunks, a chunk containing pronouns (e.g., ``he'' or ``she'') may lack speaker names, making it hard to understand. Gather allows flexible configuration of which peripheral context to include with each chunk, such as:

\begin{yaml}
- name: context_gatherer:
  type: gather
  content_key: document_text_chunk
  peripheral_chunks:
    previous:
      head:
        count: 1
        content_key: document_text_chunk
      middle:
        content_key: document_text_chunk_summary
\end{yaml}
This particular configuration includes the full content of the document's first chunk, summaries of intermediate chunks, and the current chunk itself. 
\Cref{fig:split-gather-pipeline} demonstrates different ways to render chunks. The gather operation is highly flexible in rendering contextual information, allowing for the inclusion of full chunks (as in {\em (ii)}), portions of chunks (as in {\em (i)}), or transformations (e.g., summaries) of chunks (as in {\em (iii)}). Importantly, there may be map operations between the split and gather steps---allowing for the generation of additional context (such as summaries) that can be used to augment each chunk, before downstream processing. The output adds a new attribute to each input document, containing the rendered chunk with its peripheral context, with with special tags that demarcate what is the chunk and what is peripheral context. \techreport{For additional details, see \Cref{app:gather}.}

\revision{Overall, in designing the \docetl DSL, we unified various single-document transformations (e.g., extraction, summarization) under \ttt{map} and \ttt{filter} operators, letting users express intent through prompts rather than learning multiple specialized operators. But for cross-document operations, we created distinct operators that capture specific processing patterns. For example, while \ttt{resolve} could theoretically be implemented using \ttt{equijoin}, \ttt{reduce}, and another \ttt{equijoin}, having a dedicated operator allows us to know that the user's intent is actually entity resolution, so we can better optimize the pipeline. Additionally, we distinguish \ttt{gather} from \ttt{reduce} because they serve different purposes: \ttt{reduce} performs many-to-one aggregation, whereas \ttt{gather} preserves cardinality while enriching documents with context---similar to SQL windowing functions.}

%% file: sections/rewrite.tex
\section{Rewrite Directives}
\label{sec:rewrite}

We now introduce the {\em rewrite directives} that \docetl currently supports. 
We call these {\em directives} to indicate
that they are abstract frameworks, with somewhat ambiguous semantics, that can be concretely 
instantiated by LLM agents
in a multitude of ways, as opposed to {\em rules}, which 
are more concrete, complete, and robust.
These directives are primarily designed to optimize 
the quality of outputs from \docetl pipelines 
through logical decomposition of individual operations. 
We focus on rewrite directives for map, reduce, and equijoin operators, 
with filter operators also supported through the application of map rewrite directives. We organize our rewrite directives into three main categories: data decomposition, projection synthesis, and LLM-centric improvements. 

Throughout this section, we adopt the following notation: given operators $A$ and $B$, we denote their composition as $A \to B$, where $(A \to B)(D) = B(A(D))$. For independent execution of operators, we use $A \parallel B$ to indicate that $A$ and $B$ are executed on the same input, independently. For readability, we may drop arguments---e.g., $\text{Map}_x(D)$ becomes  $\text{Map}_x$. Similarly, we omit subscripts except when the same operator appears in multiple places. We further refer to 
the text content of the document, usually stored as one of the attributes, interchangeably with the document itself, for simplicity. \revision{The arrow $\Rightarrow$ denotes a (semantic) rewrite of the operator (or operator sequence) on the left into the form on the right.}

\techreport{As mentioned previously, 
the actual instantiation and application of these directives 
are carried out by LLMs, which interpret the directives 
in the context of specific tasks and data. The benefits of each directive are also assessed by LLMs, as we can't know in advance if a directive will be helpful in a given situation. LLM agents evaluate the potential impact of each directive based on task requirements (i.e., prompts) and data characteristics, as we will discuss in \Cref{sec:optimizer}. Next, we cover each category of directives.}

\subsection{Data Decomposition}

Data decomposition is crucial 
when dealing with large documents, or when there are 
too many documents to fit in a prompt and get an accurate result for. 
We present two categories of rewrite directive here: 
{\em document chunking} and {\em multi-level aggregation}.

\subsubsection{Document Chunking (Map)}

Large documents often exceed LLM context windows or effective reasoning capabilities, leading to incomplete or inconsistent results. Our primary rewrite directive for this case, which we call the {\em split directive}, is:

\begin{equation}
\label{eq:chunking}
\begin{aligned}
\text{Map}_x & \Rightarrow \overset{\text{\textcolor{purple}{(\ref{eq:splitmap})}}}{\hspace{0.5cm}}\text{Split} \xrightarrow{\text{\textcolor{purple}{(\ref{eq:summarization})}}} \text{Gather} \xrightarrow{\text{\textcolor{purple}{(\ref{eq:filtergather})}}} \text{Map}_y \xrightarrow{\text{\textcolor{purple}{(\ref{eq:unnest})}}} \text{Reduce}
\end{aligned}
\end{equation}

Ignoring the \textcolor{purple}{purple} annotations,
this directive rewrites map to:
split the document into multiple chunks,
gather peripheral context for each chunk, 
apply a modified map operation 
per chunk, 
and  reduce the results. 
The prompt for $\text{Map}_y$ may explicitly state that only a portion of the original document is being processed. To provide more flexibility and optimization opportunities, we introduce smaller decomposition directives, for steps \textcolor{purple}{(\ref{eq:splitmap})--(\ref{eq:unnest})} above:
\begin{align}
\text{Split} &\Rightarrow \text{Map} \to \text{Split} %
\label{eq:splitmap} \\
\text{Split} \to \text{Gather} &\Rightarrow \text{Split} \to (\text{Map}_s \parallel \text{Map}_h) \to \text{Gather} %
\label{eq:summarization} \\
\text{Gather} &\Rightarrow \text{Gather} \to \text{Filter} %
\label{eq:filtergather} \\
\text{Gather} \to \text{Map} &\Rightarrow \text{Gather} \to \text{Map} \to \text{Unnest}  %
\label{eq:unnest}
\end{align}
When splitting a document, three types of context prove particularly useful: document-level metadata, hierarchical information,
and summaries of neighboring chunks. The smaller decomposition directives address these and other aspects of document processing:

\begin{itemize}[noitemsep,topsep=0pt,leftmargin=*]
\item \textbf{Document-Level Metadata Extraction (\ref{eq:splitmap}):} This directive introduces a map immediately prior to splitting, enabling the extraction of metadata relevant to {\em all} chunks. For example, when analyzing a legal contract, we might extract the contract date and parties involved from the first page, passing this information to every chunk to be rendered as part of a subsequent gather.

\item \textbf{Header Lineage Context and Summarization (\ref{eq:summarization}):} This directive introduces two independent map operations: $\text{Map}_h$ for extracting hierarchical information (e.g., headers), and $\text{Map}_s$ for generating summaries of chunks. This allows us to provide each chunk with its relevant hierarchical context (e.g., parent headers for headers in a chunk) and/or a summary of preceding content.
    
\item \textbf{Chunk Filtering (\ref{eq:filtergather}):} Not all parts of a document may be relevant for processing. This directive introduces a filter step after gathering context, allowing us to exclude irrelevant chunks.  This filter can be inferred; for instance, when processing a scientific paper, we might filter out acknowledgments or references sections if they're not pertinent to the analysis task; but they could still be used as context for other chunks if needed.

\item \textbf{Flattening Nested Results (\ref{eq:unnest}):} When processing chunks with gathered context, map might produce nested results. This directive introduces an unnest operation to flatten these results, simplifying downstream processing. For example, if each chunk produces a list of extracted entities, unnesting would flatten these lists into a single collection of entities across all chunks.
\end{itemize}

\subsubsection{Multi-Level Aggregation (Reduce)}

Large-scale aggregations can benefit from a hierarchical approach, aggregating data at a finer granularity before rolling up to the desired level. This decomposition is based on a semantic hierarchy in the data:
\begin{equation}
\label{eq:drilldown}
\text{Reduce}_{K,x} \Rightarrow \text{Reduce}_{K \cup K',y} \to \text{Reduce}_{K,z}
\end{equation}
Here $K$ is the reduce key, e.g., $K = \ttt{\{state\}}$, and $K'$ represents additional keys for finer granularity, e.g., $K' = \ttt{\{city\}}$. $y$ and $z$ are LLM-powered aggregations for the sub-reduce and final reduce operations.
For example, when summarizing voting patterns by state from social media posts, we might first aggregate data by state and city ($\text{Reduce}_{\{\text{\tt state}, \text{\tt city}\},y}$), then combine these city-level summaries to the state level ($\text{Reduce}_{\{\text{\tt state}\},z}$). This approach can capture nuances that might be lost in a single, large-scale aggregation, and allows for intermediate validation. \techreport{The effectiveness of this rewrite depends on the specific nature of the data and the aggregation task---the LLM agent must consider the appropriate granularity and design effective prompts for both aggregation steps.}

\subsection{LLM-Centric Improvements}

This category addresses unique behaviors of LLMs that can be leveraged for optimization. We present two categories of rewrite directive: {\em gleaning} and {\em duplicate resolution}.

\subsubsection{Gleaning (Map and Reduce)}

\begin{figure*}
\vspace{-30pt}
\centering
\includegraphics[width=0.9\linewidth]{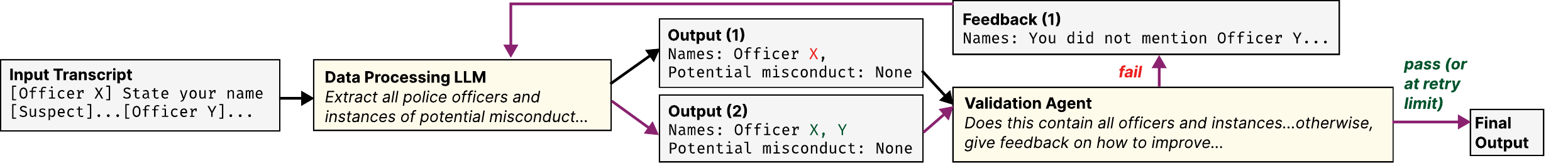}
\caption{Gleaning process with $k=1$ round of refinement. An LLM initially extracts information from an input transcript, and Officer Y is missing from the output. A validation agent (LLM-powered) identifies this omission and provides feedback. The original LLM incorporates this feedback in a second pass (shown with purple arrows), resulting in a more complete final output that includes both Officer X and Officer Y.}
\label{fig:gleaning}
\vspace{-10pt}
\end{figure*}

For this directive, we rely on the insight that when prompted with the previous inputs and outputs, and asked to improve the outputs, an LLM can iteratively refine the output.
While iterative refinement has been implemented for knowledge graph entity extraction~\cite{edge2024local}, we generalize this concept into a rewrite directive applicable to any map or reduce task. Our approach, which we call {\em gleaning}, employs separate data processing and validator LLM steps to iteratively improve output quality.
We formalize the gleaning process for map operations as:
\begin{equation}
\label{eq:gleaningrewrite}
\text{Map} \Rightarrow \text{Map} \to (\text{Map}_v \to \text{Map}_i)^{\leq k}
\end{equation}
Here, $k$ represents the maximum number of refinement iterations, $\text{Map}_v$ is a validation operation, and $\text{Map}_i$ is a refinement operation. The process works as follows:

\begin{enumerate}[leftmargin=1.5em]
    \item Init: run the original map on the input document.
    
    \item Eval: separate validator ($\text{Map}_v$) checks output based on original prompt, init's output, and a task-specific validation prompt. The validator determines if refinement is needed and describes how to improve the output, if so.
    
    \item Refine: we use a refinement map ($\text{Map}_i$) to improve the previous iteration's output based on validator feedback. Importantly, this step retains the chat history, including the original prompt, its previous response, and the validator's feedback, so it can iteratively refine.
    
    \item Iterate: repeat %
    up to $k$ times, or no further refinement is needed.
\end{enumerate}

A similar approach can be applied to reduce operations:
\begin{equation}
\label{eq:gleaningreducerewrite}
\text{Reduce} \Rightarrow \text{Reduce} \to (\text{Map}_v \to \text{Reduce}_i)^{\leq k}
\end{equation}
For reduce operations, the refinement is applied at the level of a group, not to individual documents. This enables consideration of the {\em collective} context of the grouped data.

\subsubsection{Duplicate Key Resolution (Reduce)}
A big challenge in LLM-powered data processing is that grouping,  aggregation, and summarization is difficult due to the
fact that LLM outputs are not canonicalized, and may contain many
semantic duplicates. 
To address semantic duplicates in reduce keys, especially those derived from LLM-powered operations, we introduce resolve operations:
\begin{equation}
\label{eq:resolvereduce}
\text{Reduce}_{K,x} \Rightarrow (\text{Resolve}_{k_1} \parallel \ldots \parallel \text{Resolve}_{k_m}) \to \text{Reduce}_{K,x}
\end{equation}
Where $\{k_1, \ldots, k_m\} \subseteq K$ are each a disjoint subset of keys to be resolved. Each $\text{Resolve}_{k_i}$ operation consolidates semantically equivalent values for the key $k_i$. We introduce this rewrite directive to address the inherent variability in LLM outputs: when LLMs are used to generate keys for reduce operations, they may produce semantically equivalent but syntactically different values. For example, ``New York City,'' ``NYC,'' and ``The Big Apple'' might all refer to the same entity. Without resolution, these would be treated as separate keys, leading to inaccurate aggregations.

\subsection{Projection Synthesis}
\label{sec:rewrite-projection-synthesis}

Projection synthesis strategies are inspired by projection pushdown optimizations in database systems. While selections (and selection pushdown) can also be synthesized, we did not implement this, as we found that agents are not very effective at determining whether certain data could be relevant to the query (they are overly biased by prompt wording and tend to be overly inclusive). Moreover, since an LLM-based selection is just as costly as a map, as both require an LLM call for every document, we focused on map operations that shrink the size of documents through a form of projection. \papertext{We present several instances of projection synthesis directives:} \techreport{With LLM agents, we can dynamically synthesize projections to ``push down'' based on the specific task and data at hand. However, programming LLM agents to synthesize these effectively is not straightforward, as there are potentially infinite projections that could be synthesized without necessarily improving pipeline accuracy or output quality. We present several instances of projection synthesis directives:}
\begin{align}
\text{Map}_x &\Rightarrow \text{Map}_{x_1} \to \text{Map}_{x_2} \to \cdots \to \text{Map}_{x_n} \label{eq:chainproj} \\
\text{Map}_y &\Rightarrow (\text{Map}_{y_1} \parallel \text{Map}_{y_2} \parallel \cdots \parallel \text{Map}_{y_m}) \to \text{Reduce} \label{eq:parallelproj} \\
\text{Reduce}_{K,x} &\Rightarrow \text{Map}_y \to \text{Reduce}_{K,z} \label{eq:mapreduceproj} \\
\text{Equijoin}_x &\Rightarrow (\text{Map}_{y, L} \parallel \text{Map}_{z, R}) \to \text{Equijoin}_w \label{eq:mapequijoinproj}
\end{align}

\begin{itemize}[noitemsep,topsep=0pt,leftmargin=*]
\item \textbf{Chaining} (\ref{eq:chainproj}): This directive chains simpler projections for complex map operations, useful when a map prompt contains multiple instructions. Each $\text{Map}_{x_i}$ builds on the previous result. For example, a legal document analysis could involve chained steps: extract clauses, summarize, and generate recommendations.

\item \textbf{Isolating} (\ref{eq:parallelproj}): For map operations with {\em independent} subtasks, this directive splits them into separate projections to run in parallel, followed by a reduce step. For instance, customer feedback analysis could involve \revision{isolated} projections to classify sentiment, identify features, and flag urgent issues.

\item \textbf{Pre-Aggregation} (\ref{eq:mapreduceproj}): This directive filters and projects relevant data {\em from each document} before a reduce operation, improving both efficiency and the quality of the aggregation. For example, when summarizing shipping-related feedback by product category, each detailed review could first be projected into a concise summary of shipping comments, before aggregation.

\item \textbf{Pre-Joining} (\ref{eq:mapequijoinproj}):  For complex equijoin operations, this directive preprocesses documents before joining. It is useful when direct comparison is computationally expensive---for example, matching research papers to funding opportunities could involve projecting papers to a short list of key themes and funding descriptions to criteria before joining. 
\end{itemize}

One may wonder why each operator has its own directive (e.g., map before reduce, map before equijoin). This is because the criteria for applying the directive differ by operator. For example, in pre-joining, the LLM agent evaluates factors like the sufficiency of current keys and long/large attributes. If beneficial, it generates a prompt to create a new key-value pair for a more relevant data representation. Similarly, for other operators, the agent considers operator-specific factors to determine the directive's applicability.

\techreport{Overall, our rewrite directives reflect our key insight: in complex document processing tasks, it is impossible to determine an optimal pipeline given the infinite search space, difficulty, and ambiguity. Rewrite directives provide a scaffold for systematically exploring this space, especially when coupled with opportunistic decomposition into problematic operations (as will be described in subsequent sections). The effectiveness of specific directives varies by context and is hard to predict. Finally, as a byproduct of this search process to find sufficiently accurate pipelines, we obtain interpretable pipelines, since the operators use natural language prompts.}

%% file: sections/optimizer.tex
\section{Optimizer}
\label{sec:optimizer}

Here, we detail \docetl's query planning and optimization process. Users define their pipeline in a \ttt{pipeline.yaml} file, then run \ttt{docetl build pipeline.yaml} to generate a new YAML file with an optimized pipeline.
\docetl's optimization involves two types of agents: {\em Generation agents}, which apply logical rewrite directives to create candidate plans (see ``Apply Rewrites (Agent)'' boxes in~\Cref{fig:optimizer-main}), and {\em Validation agents}, which generate custom prompts to assess the quality of these plans.
Per operation or sub-pipeline, validation agents evaluate candidate sub-plans on a data sample to select the optimal one, as shown by the green (selected) and gray (evaluated but not selected) sub-plans in~\Cref{fig:optimizer-main};
we will describe both steps next. Our framework is reminiscent of top-down approaches like Cascades~\cite{Graefe1995TheCF}, but differs in its expansion criterion (using directives) and sub-plan evaluation via LLM-based validation.  Unlike traditional cost-based optimizers, we focus on accuracy, with cost and latency constraints to be addressed in future work.

\subsection{Optimization Approach}
\label{sec:optimizer-overview}

\docetl employs a top-down optimization approach that considers both individual operations and sub-pipelines\techreport{, as outlined in Algorithm~\ref{algo:pipeline-opt} and visualized in~\Cref{fig:optimizer-main}}\papertext{, as visualized in~\Cref{fig:optimizer-main}}. We move from left to right, opting (recursively) to decompose any operations for which the accuracy is inadequate (as determined by the LLM validators). We summarize the process:

\begin{enumerate}[nosep, leftmargin=*, wide=0pt]
    \item \textbf{Pipeline Traversal and Sub-pipeline Identification}: We iterate through the pipeline from input to output (left to right). For each operation, we consider whether it, along with a suffix of the already-optimized operations to its left, forms a sub-pipeline that matches any rewrite directive. If no matching sub-pipeline is found, we treat the current operation as a single-operation sub-pipeline to optimize. For each identified sub-pipeline:
    \begin{itemize}
    \item We use the validation agent to {\em synthesize a custom validation prompt} tailored to the specific task described by the sub-pipeline.
    \item The validation agent {\em examines a sample of outputs} using this prompt to determine if there's room for improvement. If the agent concludes that the current implementation is satisfactory, we move on to the next operation without further optimization, as shown by the no-change (``NC'') paths in~\Cref{fig:optimizer-main}.
    \end{itemize} 
    \revision{\papertext{Pseudocode can be found in our technical report~\cite{shankar2024docetl}.}}
    \techreport{This process is outlined in Algorithm~\ref{algo:pipeline-opt}, and the initial validation step is shown in Algorithm~\ref{algo:subpipeline-opt} (lines 5-7).}
    
    \item \textbf{Rewrite Directive Application and Recursive Optimization}: When optimization is needed, we apply matching rewrite directives to the sub-pipeline or individual operation. As illustrated in~\Cref{fig:optimizer-main}, we explore rewrite directives from \Cref{sec:rewrite}. For each applicable directive, an LLM agent synthesizes new operations and configurations (e.g., prompts, output schemas) to match the directive. On the creation of a new operation, we immediately optimize it, recursively, before continuing with the current optimization, as shown by the nested ``Apply Rewrites'' rectangles in the figure. \techreport{This {\em opportunistic} approach allows us to explore more refined plans efficiently (Algorithm~\ref{algo:subpipeline-opt}, lines 10-11).}

    \item \textbf{Plan Evaluation and Selection}: Multiple candidate plans can arise from the rewrite directives, as depicted by the various branches in Figure~\ref{fig:optimizer-main}. We employ a two-stage evaluation process to select the best plan\techreport{, as described in Algorithm \ref{algo:plan-selection}}: First, we execute each plan on a sample of data and use the validation agent to rate the output for each document, computing an average rating per plan. We then select the top $k$ rated plans (currently set to 6) for further comparison. Next, the agent performs pairwise comparisons between these top plans, evaluating their outputs against each other. The plan with the most ``wins'' in these comparisons is selected as the optimal plan for the current sub-pipeline or operation, represented by the green boxes in~\Cref{fig:optimizer-main}. This hybrid approach balances efficiency and accuracy in plan evaluation, as pairwise comparisons are known to be ideal for assessing relative quality~\cite{parameswaran2023revisiting, liu2024aligning}, but with potentially 100+ candidate plans generated\techreport{by various rewrite directives (each rewrite can have multiple candidate plans, e.g., different parallel projections synthesized)}, comparing all pairs becomes computationally infeasible.

    \item \textbf{Pipeline Update}: We integrate the selected optimized plan into the pipeline, replacing the original operation or sub-pipeline\techreport{ (Algorithm~\ref{algo:pipeline-opt}, lines 9-12)}. 
\end{enumerate}

\techreport{\begin{algorithm}
\scriptsize
\SetAlgoLined
\KwIn{Pipeline $P$ (sequence of operators), Sample data $D$}
\KwOut{Optimized pipeline $P_{opt}$}
\BlankLine
\SetKwFunction{FOptimizePipeline}{OptimizePipeline}

\SetKwProg{Fn}{Function}{:}{}
\Fn{\FOptimizePipeline{$P, D$}}{
    $optimized \gets []$\;
    \ForEach{operation $op \in P$}{
        \uIf{$op.needsConfig$}{
            \tcp{Use LLM agent to synthesize config for new ops created by rewrite directives, including prompts, output schemas, and operator-specific parameters (e.g., reduce\_key for reduce)}
            $op.config \gets$ GenerationAgent.SynthesizeConfig($op$)\;
        }
        \uIf{$( [$\emph{suffix of} $optimized] \to op ) $ \emph{matches a rewrite directive}}{
            $subplan \gets [$matching suffix of $optimized] \to op$\;
            $optimized\_sub \gets$ OptimizeSubPipeline($subplan, D$)\;
            Replace matching suffix of $optimized$ with $optimized\_sub$\;
        }
        \Else{
            $optimized\_sub \gets$ OptimizeSubPipeline($[op], D$)\;
            Append $optimized\_sub$ to $optimized$\;
        }
    }
    \Return $optimized$\;
}

\caption{Pipeline Optimization}
\label{algo:pipeline-opt}
\end{algorithm}}

\techreport{\begin{algorithm}
\scriptsize

\SetAlgoLined
\KwIn{Sub-pipeline $SP$, Sample data $D$}
\KwOut{Optimized sub-pipeline $SP_{opt}$}
\BlankLine
\SetKwFunction{FOptimizeSubPipeline}{OptimizeSubPipeline}
\SetKwFunction{FOptimizePipeline}{OptimizePipeline}

\SetKwProg{Fn}{Function}{:}{}
\Fn{\FOptimizeSubPipeline{$SP, D$}}{
    \uIf{$SP$ \emph{does not match any rewrite directive}}{
        \Return $SP$\;
    }
    Execute $SP$ on $D$ to get outputs\;
    \tcp{Synthesize a prompt for validating sub-pipeline output}
    $V \gets$ ValidationAgent.SynthesizeValidatorPrompt($D$, outputs, $SP$)\;
    \uIf{ValidationAgent.Validate(outputs, $V$) \emph{is satisfactory}}{
        \Return $SP$\;
    }
    $candidate\_plans \gets []$\;
    \ForEach{\emph{directive} $R \in$ \emph{applicable rewrite directives for} $SP$}{
        \tcp{R applied to SP generates a mix of old and new ops}
        $rewritten\_ops \gets$ $R$ applied to $SP$\;
        $plan \gets$ \FOptimizePipeline{$rewritten\_ops, D$}\;
        Append $plan$ to $candidate\_plans$\;
    }
    \Return $\text{PlanSelection}(candidate\_plans, V, D, k)$\;
}

\caption{Sub-pipeline Optimization}
\label{algo:subpipeline-opt}
\end{algorithm}}

\techreport{\begin{algorithm}
\scriptsize
\SetAlgoLined
\KwIn{Candidate plans $C$, Validation prompt $V$, Sample data $D$, Number of top plans to compare $k$}
\KwOut{Best plan $best\_plan$}
\BlankLine
\ForEach{\emph{plan} $p \in C$}{
    Execute $p$ on each sample in $D$\;
    Use ValidationAgent to rate outputs on a scale of 1 (very bad) to 4 (no identified improvements) according to $V$\;
    Compute average score for $p$ across samples\;
}
Select top $k$ plans based on average scores\;
\BlankLine
\ForEach{\emph{pair of plans} $(p_i, p_j)$ \emph{in top $k$ plans}}{
    Perform pairwise comparison using ValidationAgent and $V$\;
    Update comparison scores for $p_i$ and $p_j$\;
}
\BlankLine
\Return plan with highest comparison score\;
\caption{Plan Selection}
\label{algo:plan-selection}
\end{algorithm}}

To execute candidate plans (so we can compare their outputs), we sample data based on document size (larger documents have higher selection probability). As we optimize each sub-pipeline, we track its selectivity ratio (output documents / input documents) and use these ratios to adjust sample sizes for later operations. For example, if the first two operations have selectivities of 0.5 and 0.3, we increase the initial sample size by $(1 / 0.5 / 0.3) \approx 6.67$ when optimizing the third operation. This ensures sufficient data for optimization even after selective operations. However, sample documents may not fully represent the complete dataset; e.g., if the sampled documents fit within LLM context limits but some documents in the full dataset exceed them, we may encounter errors during full execution. We are developing methods to adapt plans accordingly during pipeline execution time.

\techreport{Our overall approach lends itself to a rich space of pipeline optimization techniques with operator reordering and operator fusion. While we have not implemented any in the current release of \docetl, we are actively exploring this area for future improvements.}

\subsection{Agent and System Implementation}
\label{sec:agent-arch}

\techreport{Here, we outline our novel agent-based architecture for generation and validation. While a comprehensive analysis of our architectures is beyond the scope of this paper, we focus on critical aspects that significantly impact system performance and effectiveness.

\subsubsection{Generation Agents} Generation agents are responsible for applying rewrite directives to create diverse candidate plans. When presented with a directive, these agents synthesize one or more appropriate operation configurations. These configurations encompass both logical and physical choices. Logical choices include prompts, output schemas, and reduce keys, while physical choices involve parameters such as chunk sizes for document splitting and batch sizes for document reduction. The generation agent also evaluates the applicability of rewrite directives in specific contexts. For instance, the agent might determine that applying the split-map directive (\Cref{eq:splitmap}) is not beneficial if there's no valuable document-level metadata to leverage when processing individual chunks.

For certain parameter choices, particularly those related to physical implementation, LLMs may not be well-suited to determine optimal values. For example, how would an LLM know the ideal number of documents to summarize together in a batch as part of a reduce operation? In these cases, we use heuristics to generate a range of plausible parameter values, such as different batch sizes for a reduce operation, and then compare the results of these plans to determine the most effective parameter choice for the given operation and context. 

Here, we detail three examples of our generation agent's approach for parameter selection: 

\topic{Chunk Sizes} Our chunking approach explores five sizes ranging from 15\% to 75\% of the LLM's context limit, uniformly sampled. We also explore chunk sizes based on percentages of the {\em average document length}; similarly six sizes ranging from 15\% to 100\%, uniformly sampled. also For each chunk size, we generate a set of gather configurations to retain relevant context from surrounding chunks. The creation of these gather configurations is based on the ratio of chunk size to document size.

    We begin with three base configurations of gather operations for each chunk size: no context, one previous chunk, and one previous plus one next chunk. We then expand this set based on the document-to-chunk size ratio. For larger ratios (indicating smaller chunks relative to the document size), we generate configurations with more peripheral context. We use a square root function to control the growth of peripheral context as the document-to-chunk ratio increases, preventing excessive context that could overwhelm the model. The choice of square root is based on empirical observations that the benefit of additional context tends to diminish more drastically as more context is added---a detailed evaluation is left for future work. For example, if the document is significantly larger than the chunk size, our expanded set might include configurations with up to 5 previous chunks and 2 next chunks. Conversely, for ratios closer to 1 (where chunk size approaches document size), our set comprises only the base configurations.

    This basic approach is a first attempt at systematically exploring various chunking and gathering strategies. We are currently developing a taxonomy of LLM-powered data processing tasks to further refine this process. Our goal is to eventually use task classification to guide the generation of more tailored chunk sizes and gather configurations, recognizing that optimal settings may vary significantly depending on the specific task at hand.

\topic{Batch Sizes} For reduce operations, optimal batch sizes (i.e., the number of documents aggregated at once, in a single prompt) are not obvious and require experimentation. Our agent tests sizes from 20\%, 40\%, 60\%, 75\%, to 90\% of the maximum input fitting the LLM's context window, generating and evaluating multiple fold prompts for each. Our evaluations reveal task-dependent optimal batch sizes, highlighting the need for further research in this area---some tasks perform best with the smallest batch size (e.g., extracting distinct names), while others peak at a middle batch size, as shown in \Cref{sec:evaluation}.
    
 \topic{Blocking Keys and Rules} Resolve and equijoin operators involve pairwise comparisons between entities or records, leading to quadratic complexity in LLM calls. To mitigate this, a common technique is to use {\em blocking} to filter the number of pairs~\cite{eroverview}. \docetl offers two blocking approaches: embedding-based and code-based. Embedding-based blocking leverages an embedding model (default: OpenAI's text-embedding-3-small) to generate vector representations for each document or subset of key-value pairs in a document (i.e., {\em blocking keys}). We compute cosine similarities between these embeddings and only consider pairs whose similarity exceeds a specified threshold for full LLM-based comparison. Code-based blocking allows custom Python expressions to be specified as filters. While blocking keys  and code-based blocking rules can be directly constructed by the generation agents, we employ a different approach for determining the embedding threshold. Instead of asking an LLM to arbitrarily come up with a similarity threshold, we empirically determine it: first, we sample hundreds of pairs that are likely to be duplicates based on their embedding similarity. We then execute the comparison prompt on these pairs to identify the true duplicates. Finally, we select the threshold that achieves 95\% recall in duplicate identification.

\subsubsection{Validation Agents} Validation agents assess sub-pipeline effectiveness through a structured validation and comparison process. For each operator, they synthesize validation criteria focused on concrete properties like accuracy (correctness of extracted information), precision (avoiding hallucinated content), and recall (completeness of extracted information), rather than relying on the operation's prompt instructions. These criteria are formulated as explicit tests that can be systematically checked against operation outputs.

To evaluate operation outputs, the agents first process a sample of data and assess the outputs against the synthesized validation criteria. This assessment determines whether further optimization is needed based on concrete failures rather than subjective assessment. When comparing candidate plans, the agents employ a two-stage approach: first rating each plan's outputs on a scale from 1 (very bad) to 5 (excellent) based on how well they meet the validation criteria, then performing detailed pairwise comparisons between the top-$k$ rated plans for a more nuanced quality assessment, as described in Algorithm \ref{algo:plan-selection}. We currently set $k=6$ to balance thorough evaluation with computational efficiency, though we leave a more systematic parameter selection strategy for future work.

This structured approach to validation enables us to identify specific failure modes and guide the optimization process toward concrete improvements, similar to how traditional software testing isolates bugs through specific test cases. The validation criteria serve as a consistent benchmark across different pipeline variants, allowing for reliable comparison of alternative plans.

\subsubsection{Implementation Details} \docetl leverages GPT-4o (OpenAI) as the default LLM for both generation and validation agents, but this can be changed by the user.  Both generation and validation agents consider a variety of inputs in their prompts, including user-defined operation prompts, sample operation input data, and, when relevant (i.e., for evaluation), sample operation output data. Often, including {\em all} of this data in a single prompt exceeds the LLM's context limits. When this happens, we have to remove information from the prompt. We prioritize keeping the following types of information:

\begin{enumerate}[nosep, leftmargin=*, wide=0pt]
\item \emph{Output Schema Attributes}: These are given the highest priority, with all tokens included---which is feasible because LLM output limits are typically much smaller than prompt (i.e., input) limits.
    
\item \emph{Prompt-Referenced Attributes}: Of next priority is input attributes explicitly referenced in the prompt template, ensuring the LLM has access to all task-critical information.
    
\item \emph{Remaining Input Attributes}: For any additional attributes in the input document(s), we implement a middle truncation strategy. This method preserves both the initial and final portions of the content, which often encapsulate key information, while judiciously truncating the middle sections as necessary.
\end{enumerate} 

To optimize performance and resource utilization, we cache all sub-pipeline outputs. The engine and optimizer are implemented in approximately 16K lines of Python, with 2K lines in Rust for efficient resolve and equijoin execution. While blocking rules are defined in Python, so they can be easily generated by LLMs, we implement common patterns (like containment and normalized string matching) in Rust for better performance. Structured outputs for LLM calls are handled by the tool calling functionality; users can use any LLM in \docetl pipelines that supports tool calling (e.g., OpenAI, Claude, Gemini, Llama 3.1, and more).}

%% file: sections/new_eval.tex
\section{Evaluation}
\label{sec:evaluation}

The primary goal in our evaluation is show that \docetl's rewrite directives and optimization framework dramatically enhance our ability to automatically analyze complex documents---all with no training labels or developer intervention needed. While finding optimal plans is impossible, we demonstrate that \docetl's approach of systematically decomposing tasks and documents to explore a search space of processing strategies yields plans that are sufficiently accurate. In comparison, baseline approaches often achieve such poor accuracy (< 40\%) that they are impractical for real use. 
 
 {\bf \em Overall, we find that \docetl's plans yield 21\% to 80\% 
improvements in task-specific accuracy metrics such as precision, recall, and F1 score.} We first consider three complex document processing tasks: legal contract analysis, declassified article analysis, and video game review analysis \revision{(\Cref{sec:evaluation-legal,sec:evaluation-declassified,sec:evaluation-games})}. These tasks represent different challenges: extracting structured information embedded within the semantic content of unstructured data, resolving entities and summarizing their information across documents, and reasoning about temporal consistency across long documents.
\revision{For the legal contract analysis (\Cref{sec:evaluation-legal}), we compare against both recent LLM-powered systems (LOTUS~\cite{patel2024lotus}, Palimpzest~\cite{liu2024declarative}, and Aryn~\cite{anderson2024designllmpoweredunstructuredanalytics}) and traditional NLP baselines using spaCy~\cite{spacy} or NLTK~\cite{bird2009natural}. For the video game review  (\Cref{sec:evaluation-games}) and declassified article  (\Cref{sec:evaluation-declassified}) tasks, we compare only against non-LLM baselines, as LOTUS, Palimpzest, and Aryn lack support for entity resolution and documents exceeding LLM context windows. For each task, our evaluation includes both task-specific metrics (customized variations of precision and recall) as well as a hallucination rate to measure factual consistency. Then, we evaluate \docetl on the challenging Biodex text classification task (LOTUS's only task with sub-70\% accuracy), where our optimized pipeline achieves {\em \bf 33 to 80\% improvements} in rank precision (\Cref{sec:evaluation-biodex}). \papertext{We conclude with case studies examining \docetl's application in real-world police misconduct identification, the effectiveness of LLM agent rewrites, and insights from user adoption (\Cref{sec:evaluation-vldbcasestudies}).}\techreport{We conclude with case studies examining \docetl's application in real-world police misconduct identification, the effectiveness of LLM agent rewrites, and insights from user adoption (\Cref{sec:evaluation-misconduct,sec:evaluation-legal-casestudy,sec:eval-adoption}).}}

For all pipelines, 
we use the gpt-4o-mini model from OpenAI, 
and we run the experiments on a 2021 Macbook Pro with an M1 chip. \revision{The \docetl optimizer uses gpt-4o-mini, except in the Biodex task in \Cref{sec:evaluation-biodex}, where we use gpt-4o.}
\papertext{Additional implementation details can be found in our
technical report~\cite{shankar2024docetl}.}

\subsection{Legal Contract Analysis}
\label{sec:evaluation-legal}

The Contract Understanding Atticus Dataset (CUAD)~\cite{hendrycks2021cuad},
includes 510 legal contracts 
with expert-labeled annotations 
across 41 categories of  clauses, ranging from basic information (e.g., Document Name, Parties) to complex concepts (e.g., Most Favored Nation, IP Ownership, Post-Termination Services). 
The task is to extract text spans 
for each relevant clause type from each contract; 
not all contracts contain all types of clauses.

We evaluate on the first 50 contracts, comparing extractions against ground truth. 
An extraction is considered correct if 
{\em (i)} the clause type matches, 
and {\em (ii)} the extracted text span's Jaccard similarity with the ground truth span $>0.15$. This threshold accommodates variation in LLM outputs while ensuring the model has correctly identified the clause's location;
it is set fairly low because we provide no training examples, 
so the LLM does not know how much to extract---but 
large enough to ensure some match. 
We set other values for this and found the comparisons
to be similar. 
We measure precision, recall, F1, and \revision{hallucination rate (proportion of extracted clauses not matching our 41 predefined categories)}.

\subsubsection{Implementations}

We have \revision{five} baselines:

\begin{enumerate}[nosep, leftmargin=*, wide=0pt]
\item {\bf \docetl Baseline:} Our unoptimized pipeline consists of a single map with a prompt to extract all relevant clauses, given one-sentence descriptions of the 41 clause types. The output schema specifies a list of objects with \texttt{clause\_type} and \texttt{text\_span} keys. \papertext{The pipeline code is given in our technical report~\cite{shankar2024docetl}.}

\begin{yaml}
- name: extract_relevant_clauses
  type: map
  output:
    schema:
      misconduct: "list[{clause_type: str, text_span: str}]"
  prompt: |
    Given the following contract document:
    {{ input.document }}
    Extract the text spans (if they exist) for each of the following categories:
    1. Document Name: The name of the contract
    2. Parties: The two or more parties who signed the contract
    3. Agreement Date: The date of the contract
    4. Effective Date: The date when the contract is effective
    ...37 more...
\end{yaml}

 \item {\bf LOTUS Baseline:} We implement a pipeline using LOTUS's \ttt{sem\_map} operator with the same prompt as \docetl's map operation, plus additional output structuring instructions since LOTUS does not support explicit output schema definitions. \techreport{The LOTUS pipeline output is a string that we parse into JSON for evaluation. While some outputs required re-running to obtain parseable JSON, we report costs for a single run to maintain fair comparison. {\bf We expect the accuracies for the LOTUS baseline to be similar to the \docetl baseline}, as the LLM calls are mostly the same; the only differences arise from discrepancies in the system prompts (part of the \docetl and LOTUS codebases; not exposed to the user), as well as the extra instruction in the LOTUS prompt to output a JSON-formatted answer matching the intended schema (which contains examples of some of the clause types).}

\item {\bf Palimpzest Baseline:} We implement the extraction using Palimpzest's \texttt{convert} operator. In Palimpzest, rather than writing prompts directly, users provide schema descriptions from which the system generates prompts. We provided our clause type descriptions in the description of the schema.

\item {\bf Non-LLM Baseline:} \revision{We write a program, using the spaCy library~\cite{spacy}, to loop over all clause types and extract the most semantically similar sentence (above a threshold of 0.9). We use spaCy's sentence splitter and embedding model, tok2vec.}
\item {\bf Aryn Baseline:} \revision{We implement extraction using Aryn's \ttt{llm\_query} operation with the same prompt as our LOTUS baseline, and the same output normalization procedure used with LOTUS to handle parsing errors and format inconsistencies.}

\item {\bf \docetl's Optimized Plan:} \docetl's optimizer transforms the 
single map operation into an isolated projection 
decomposition with 21 \revision{independent} map operations, each extracting 1-3 semantically related spans (e.g., grouping agreement and effective date extractions),
followed by a reduce to combine all extracted clauses. 
Notably, the optimizer chose 
isolated projection (directive \ref{eq:parallelproj}) over document chunking, suggesting that LLMs excel at focused extraction of small amounts of information even from lengthy documents.
\end{enumerate}

\subsubsection{Results}
The results are shown in~\Cref{tab:legal-results}. \docetl's optimized plan performs significantly better than all baselines, achieving a {\bf \revision{21.4\%} improvement in F1 over LOTUS}, the next best LLM-based plan, and a {\bf 67\% improvement in recall} over the unoptimized \docetl pipeline\revision{---with no hallucinations. LOTUS, Aryn, and the unoptimized \docetl pipelines achieve similar scores and hallucination rates (6.9-7.3\%). The non-LLM baseline achives much lower scores than the LLM-based methods, as well as longer text spans---because spans are forced to be at sentence-level granularity, which could be longer than necessary for short clauses like ``document name'' or ``agreement date.''} Interestingly, Palimpzest's optimizer selected a code-based plan rather than an LLM-based one for this task\revision{---perhaps explaining its lower score}. \techreport{Palimpzest's lower performance on this specific task may be due to difficulties in configuring its schema-only approach.}

While the optimized pipeline's cost 
and runtime are higher (\Cref{tab:runtime-cost}), 
we prioritize accuracy, 
which often requires increased computational costs. 
The higher runtime and cost 
stems from the increased number of LLM calls 
in the \revision{new} map operations, 
plus an additional reduce operation to combine their results. Further parallelism could help reduce the runtimes
further, but this is not our focus.
Costs will decrease as LLM pricing continues to fall---they have fallen by 1000$\times$ in 3 years, with a predicted drop of 10$\times$ per year~\cite{llmflation}---and they become negligible when using open-source models. 
The optimization cost is only \$1.58 (using gpt-4o-mini\techreport{ for the optimizer's agents}) and does not increase with dataset size, 
as it is done on a sample.

\begin{table}[t]
\caption{Legal Contract Analysis Results.}
\label{tab:legal-results}
\footnotesize
\centering
\vspace{-10pt}
\begin{tabular}{l>{\centering\arraybackslash}p{0.75cm}>{\centering\arraybackslash}p{0.75cm}c>{\centering\arraybackslash}p{0.75cm}>{\centering\arraybackslash}p{1.5cm}}
\toprule
System & Avg Precision & Avg Recall & Avg F1 & \revision{Avg \# Chars} & \revision{Avg Hallucination Rate} \\
\midrule
\docetl (Opt.) &  \revision{0.401} & \revision{\bf 0.719} & \revision{{\bf 0.477}} & \revision{162.60} & \revision{{\bf 0.000}} \\
\docetl (Unopt.) & \revision{0.341} & \revision{0.430} & \revision{0.379} & \revision{49.35} & \revision{0.072} \\
LOTUS & \revision{0.402} & \revision{0.471} & \revision{0.393} & \revision{46.301} & \revision{0.073} \\
Palimpzest & 0.059 & 0.013 & 0.022 & \revision{35.10} & \revision{{\bf 0.000}} \\
\revision{Aryn} & \revision{{\bf 0.450}} & \revision{0.370} & \revision{0.352} & \revision{49.56} & \revision{0.069} \\
\revision{Non-LLM} & \revision{0.224} & \revision{0.219} & \revision{0.190} & \revision{212.73} & \revision{{\bf 0.000}} \\
\bottomrule
\end{tabular}
\vspace{-10pt}
\end{table}

\begin{table}[t]
\caption{Runtime and Cost Analysis for Legal Task. \techreport{N/A means not available, because the pipeline or system does not have an optimizer. }Palimpzest runtime is single-threaded and includes optimization time.}
\label{tab:runtime-cost}
\vspace{-10pt}
\footnotesize
\centering
\begin{tabular}{lccc}
\toprule
System & Runtime (s) & Cost (\$) & Optimizer Cost (\$) \\
\midrule
\docetl (Opt.) & 180.30 & 1.46 & 1.58 \\
\docetl (Unopt.) & 23.43 & 0.08 & N/A \\
LOTUS & 28.12 & 0.07 & N/A \\
Palimpzest & 84.07 & Unknown$^*$ & Unknown$^*$ \\
\revision{Aryn} & \revision{52.53} & \revision{Unknown$^*$} & \revision{N/A} \\
\revision{Non-LLM} & \revision{217.99} & \revision{0.00} & \revision{N/A} \\
\bottomrule
\end{tabular}
{\par\raggedright \footnotesize $^*$Costs are not reported by the system.\par}
\vspace{-10pt}
\end{table}

\subsection{Game Review Analysis}
\label{sec:evaluation-games}

\begin{table}[t]
\caption{Game Review Analysis Results}
\label{tab:game-review-results}
\vspace{-10pt}
\footnotesize
\centering
\begin{tabular}{lp{1cm}p{1cm}p{1cm}}
\toprule
Metric  & \docetl (Unopt.) & \docetl (Opt.) & \revision{Non-LLM} \\
\midrule
Hallucination Rate (lower is better) & 0.465 & \textbf{0.312} & \revision{N/A} \\
Sentiment Accuracy (higher is better) & \textbf{0.664} & 0.650 & \revision{0.605} \\
Kendall's Tau (higher is better) & 0.470 & \textbf{0.631} & \revision{N/A} \\
\bottomrule
\end{tabular}
\vspace{-10pt}
\end{table}

We evaluate \docetl on temporal analysis of video game reviews from Steam\footnote{\url{https://www.kaggle.com/datasets/najzeko/steam-reviews-2021}}. 
For each of 10 popular games\techreport{ (randomly sampled from the 100 games with the most reviews)}, we create a document 
with 300 customer reviews with timestamps 
(but omit their ratings). 
Each document comprises concatenated 
reviews in no particular order, with lengths exceeding standard LLM context windows. The task is to identify 10 positive and 10 negative reviews per game, with their review IDs, and present these in chronological order.
We evaluate the pipelines 
on: {\em (i)} hallucination rate, 
or the fraction of extracted review IDs 
that do not appear in the source, 
{\em (ii)} sentiment accuracy: whether 
the identified review sentiment matches the user's rating, 
computed only for non-hallucinated reviews, 
and {\em (iii)} Kendall's Tau correlation 
of the timestamp ordering, which measures 
how well the reviews are chronologically ordered.

\subsubsection{Implementations}
Since the documents exceed context limits\techreport{ and require temporal reasoning}, 
we do not compare against existing \revision{LLM-based} systems, which do not support documents beyond context windows. 
Our baseline \docetl pipeline consists of a single map to extract \texttt{positive\_reviews} and \texttt{negative\_reviews}\techreport{ (both list types)}, with documents truncated from the middle to fit the context window---effectively randomly sampling reviews\techreport{ from each game's corpus}. \papertext{The exact pipeline can be found in our techical report~\cite{shankar2024docetl}.} %
The operation looks like the following:

\begin{yaml}
- name: get_reviews
  type: map
  output:
      schema:
        positive_reviews: "list[{review_id: str, timestamp: str, review_summary: str}]"
        negative_reviews: "list[{review_id: str, timestamp: str, review_summary: str}]"
  prompt: |
     Given the following reviews for the game {{ input.app_name }}, analyze them and select 10 positive and 10 negative reviews that are evenly distributed across time: {{ input.concatenated_reviews }}
     Return two lists:
     - positive_reviews: List of 10 positive reviews, sorted by timestamp
     - negative_reviews: List of 10 negative reviews, sorted by timestamp
     Each returned review object should contain the review ID, timestamp and a summary of the review.
\end{yaml}

\docetl's optimizer transforms this pipeline into: (a)
A {\em split} operation that chunks input by token count (104,652 tokens per chunk): no gather operation
(b) Two {\em map} operations per chunk---one each for positive/negative reviews---each incorporating one round of {\em gleaning} (directive \ref{eq:gleaningrewrite}) to ensure that the reviews are valid
(c) A {\em reduce} operation to combine the positive and negative reviews from the chunks and present them in chronological order\techreport{, matching the original \ttt{get\_reviews} operation's output schema}.
\revision{We added a non-LLM baseline that extracts reviews via regex, classifies sentiment with NLTK and VADER~\cite{bird2009natural, hutto2014vader}, and selects the first 10 positive and negative reviews. Since this baseline only performs classification (i.e., it is not a generative model), hallucination rate and Kendall's Tau metrics don't apply.}

\subsubsection{Results}

As shown in \Cref{tab:game-review-results}, we observe a {\bf 32.9\% reduction in hallucination rate}\techreport{ (from 46.5\% to 31.2\%)}, demonstrating more reliable review extraction. Sentiment accuracy remained stable (66.4\% vs 65.0\%), while Kendall's Tau improved by {\bf 34.3\%}, indicating better temporal ordering. \revision{Both LLM-based approaches outperform the non-LLM baseline in sentiment accuracy, despite having to handle complex additional tasks beyond simple sentiment classification.}

The optimized pipeline costs \$1.48 (173.63s runtime) versus the baseline's \$0.12 (29.27s). However, the baseline achieves this by truncating data to fit LLM context limits. With full data processing, the baseline would cost \$0.28\techreport{, making the optimized pipeline $5.3\times$ more expensive---still less than the $10\times$ cost gap between gpt-4o and gpt-4o-mini models}. This cost increase is justified by the improved temporal reasoning accuracy, and is due to steps like gleaning (which doubles operation cost); however, the gleaning validator consistently flagged temporal issues; with feedback like ``The ... reviews are not sorted correctly by timestamp; they should be organized chronologically.''  
The optimization cost was \$6.60; however, this is a one-time cost. \revision{The non-LLM baseline had a runtime of 15.89 seconds.} 

\subsection{Declassified Article Analysis}
\label{sec:evaluation-declassified}

We evaluate \docetl's effectiveness 
on resolve and reduce tasks 
using 733 paranormal case files from The Black Vault, 
a repository of declassified international government documents,
averaging 700 words each.
Each article documents a reported paranormal event 
with details such as location and witness accounts. We scraped articles from their website and used Azure Document Intelligence to convert all PDF attachments to text\techreport{, and provide this data for transparency at \url{https://osf.io/9xsbq}}.
Our task is to determine the distinct locations for 
each type of paranormal event\techreport{ (e.g., all cities and regions where UFO sightings were reported)}. 
The task involves two challenges: {\em (i)} standardizing event types across articles, 
and {\em (ii)} extracting and aggregating 
location mentions across articles for each event type. 

We evaluated precision of extracted locations 
by first programatically verifying their presence 
in the source text 
and attempting to geocode them using the Nominatim API, based on OpenStreetMap. 
\revision{We also measured hallucination rate---a subset of precision---defined as the proportion of locations that don't exist in the source text.}
For locations in the text that could not be geocoded
(e.g., specific rivers or mountain ranges), we performed manual verification.

\subsubsection{Implementations}
We consider \revision{4} pipelines. 
We only consider one \revision{LLM-powered} baseline, written in \docetl, 
as other systems don't support resolve. This pipeline consists of: {\em (i)} a map to extract event type (e.g., ``humanoid sighting'') per article, 
and {\em (ii)} a reduce to collect distinct locations  
across all articles of each event type. \docetl's optimizer modified this pipeline in two ways. 
First, it synthesized a resolve between map 
and reduce to standardize event types (directive \ref{eq:resolvereduce}). 
Second, it optimized reduce by determining a fold batch size (41) to 
process document batches\techreport{, synthesizing the corresponding fold prompts}. \techreport{The optimized pipeline consists of: {\em (i)} 
a map (as before), {\em (ii)} a resolve 
to standardize event types (e.g., variations of ``UFO sighting''), 
and {\em (iii)} a reduce (as before), but using a batched fold,
with batch size 41.} To isolate the impact of the optimized reduce operation, we also evaluate the a version of this pipeline {\em (+resolve only)}, 
which uses the original reduce operation without batched folding. \revision{Our 4th pipeline represents a non-LLM baseline that extracts location (LOC) entities from article text using spaCy's \ttt{en\_core\_web\_lg} model~\cite{spacy}. This script processes the resolved results from \docetl's optimized pipeline to establish a comparison point for location precision and recall.}

\subsubsection{Results}

\begin{table}[t]
\vspace{-10pt}
\caption{Declassified Article Analysis Results. Location metrics for baseline are N/A as its 233 distinct event types (mostly singleton categories) make meaningful location aggregation impossible.}
\label{tab:result-declassified}
\centering
\footnotesize
\begin{tabular}{lp{1cm}p{1.5cm}p{1cm}p{1cm}}
\toprule
Metric & \docetl (Unopt.)  & \docetl (+Resolve Only) & \docetl (Opt.) & \revision{Non-LLM} \\
\midrule
Location Precision & N/A  & 0.994 & {\bf 1.000} & \revision{0.6812} \\
Location Recall & N/A & 298 & {\bf 435} & \revision{{\bf 435}}  \\
Distinct Event Types & 164 & 83 & 83  & \revision{N/A} \\
\revision{Hallucination Rate} & N/A  & \revision{0.01} & \revision{0.01} & \revision{0.00} \\
\bottomrule
\end{tabular}
\vspace{-15pt}
\end{table}

As shown in \Cref{tab:result-declassified}, the baseline \docetl pipeline extracts 233 distinct event types with many semantic duplicates (e.g., ``UFO Sighting'', ``Category: UFO Sighting'', ``Event Type: UFO Sighting''), making location aggregation impractical as most event types contain only one article. Adding resolve enables meaningful aggregation by consolidating to 83 event types. The +resolve only pipeline extracts 298 locations with 99.4\% precision, and the optimized pipeline further improves this to 100\% precision while extracting 435 locations {\bf (46\% higher recall)}. \revision{The non-LLM baseline matches the optimized pipeline's recall but with substantially lower precision (68.12\% vs. 100\%), highlighting the LLM's superior ability to {\em accurately} identify relevant locations in context. All systems exhibit low hallucination rates.} This improved recall arises because batched folding allows the LLM to incrementally process and track distinct locations, rather than attempting to process all documents at once, where important details may be lost due to LLM context window overload~\cite{levy2024same, liu2024lost}.

\techreport{\topic{Cost Analysis} } The resolve-only pipeline cost \$1.16 (307.36s), while the optimized version cost \$1.84 (\$1.34 + \$0.50 for optimization; 625.64s). The optimized pipeline's longer runtime results from multiple LLM calls per event type during folding, versus one call in the resolve-only version. \revision{The non-LLM baseline ran in 158.85s.} \techreport{For all operations, and the optimizer agents, we used gpt-4o-mini.}

\subsection{Biomedical Classification}
\label{sec:evaluation-biodex}

We evaluate \docetl on the challenging Biodex biomedical drug reaction classification task from the LOTUS paper~\cite{patel2024lotus}. For each of 250 biomedical papers, the task involves identifying which of 24,300 adverse drug reactions (from the MedDRA list) are discussed. Performance is measured using rank-precision@k (RP@k), evaluating both accuracy and ranking of identified reactions. A higher score indicates that true positive reactions appear earlier in the list. \revision{We also evaluate the hallucination rate, measuring the proportion of identified reactions that are not present in the drug reaction list.}

\subsubsection{Implementations}

We compare against LOTUS using numbers from their preprint and our reimplementation of their pipeline using the same models (gpt-4o-mini for LLM calls, text-embedding-3-small for embeddings) for fair comparison. \papertext{Note that for joins, their codebase as of November 2024 deviates
from their preprint by invoking an LLM for
each tuple pair, which would require over 
6 million LLM calls, which is prohibitive.
We describe how we reimplemented their pipeline in our technical report~\cite{shankar2024docetl}.}

\techreport{\topic{LOTUS Baseline} We implemented their {\em map-search-filter} pattern as a pandas dataframe accessor: first extracting reactions from each article via a map operation, then using similarity search to find candidate MedDRA labels, and finally filtering these candidates with an LLM to verify matches.

Our implementation follows their described pipeline with some necessary deviations. First, a map operation extracts drug reactions from each article (one LLM call per article). Then, for each article's extracted reactions, we find the 49 nearest neighbors in embedding space among the MedDRA labels---we chose k=49 to match our target LLM call budget ($\sim12k$, equivalent to the number of LLM calls executed by \docetl). We use exact cosine similarity computation via NumPy instead of a FAISS approximate nearest neighbor index, which may slightly increase runtime but provides more precise results---though this overhead is minimal given we only search through O(10k) vectors per article on an M1 Mac. 

To evaluate rank-precision, we post-processed the pipeline outputs into ranked lists. For each article, we ranked the matching labels based on their cosine similarity scores (obtained during the similarity search phase), taking the top k labels for RP@k computation.

\topic{\docetl Implementation}}
In \docetl, we implement this 
task as an equijoin between articles and MedDRA labels, 
using a comparison prompt that asks ``Can the following condition be found in the article?'' \techreport{The prompt includes both the article text and the label, as well as an indicator of whether the condition text appears as a substring of the article (which is possible in Jinja templating).} We don't evaluate an unoptimized version due to the impractical number of LLM calls required (over 6 million).
\docetl optimized this into a map-equijoin pipeline, where the map extracts medical conditions per article\techreport{(with a prompt designed for medical text but without demonstrations or examples)}, and the equijoin uses synthesized blocking rules including an embedding similarity threshold of 0.5253 and a requirement that all words in the reaction label appear in the article text. \techreport{Finally, we add a reduce operation that asks the LLM to rank the identified labels for each article from most to least confident, enabling evaluation of ranking quality. We did not apply \docetl's reduce operator optimizations to this ranking step.}

\revision{We also include a simple non-LLM baseline that identifies candidate labels by checking for exact substrings and ranks them by length. Given the large number of required comparisons, we opted for a keyword baseline over more complex NLP libraries (e.g., NTLK).} \papertext{All outputs were post-processed into ranked lists (based on length of match) for evaluation. Additional implementation details can be found in our technical report~\cite{shankar2024docetl}.}

\subsubsection{Results}

\techreport{The two pipelines differ in how they rank identified labels for each article. \docetl uses a reduce operation that asks the LLM to rank labels from most to least confident, while for LOTUS we use semantic similarity scores from the search phase, as their semantic aggregation operator operates on entire dataframes rather than groups, making LLM-based ranking per article outside the scope of their implementation.}

\begin{table}[t]
\vspace{-10pt}
\caption{Biomedical Classification Results. Since most articles have fewer than 25 relevant labels, RP@25 effectively measures recall rather than ranking quality.}
\label{tab:biodex-results}
\vspace{-10pt}
\centering
\footnotesize
\begin{tabular}{p{3.5cm}cccc}
\toprule
System & RP@5 & RP@10 & RP@25 & \revision{Hallucination} \\
\midrule
DocETL & \textbf{0.281} & \textbf{0.313} & \textbf{0.371} & \revision{0.001} \\
LOTUS (Our reimplementation \revision{of map-search-filter}) & 0.213 & 0.207 & 0.206 & \revision{0.000} \\
LOTUS (Reported) & 0.241 & 0.258 & N/A & \revision{0.000} \\
\revision{Non-LLM Baseline} & \revision{0.106} & \revision{0.158} & \revision{0.262} & \revision{0.000} \\
\bottomrule
\end{tabular}
\vspace{-20pt}
\end{table}

At RP@25, which effectively measures recall since articles contain fewer than 25 labels in the ground truth, \docetl shows an {\bf 80\% improvement} over reimplemented LOTUS. For RP@5 and RP@10, \docetl shows {\bf 33\% and 50\% improvements}, respectively. \revision{The non-LLM baseline achieves lower RP@5 and RP@10 scores than the LLM-based methods, but a competitive RP@25 (still worse than the plan \docetl found).} The improvement in recall likely stems from \docetl's synthesized blocking rules: while LOTUS relies purely on embedding similarity to identify candidate reactions, {\bf \em \docetl's rule requiring all words in the reaction label to appear in the article text may surface reactions that have low embedding similarity scores.} \revision{In terms of hallucination rate, all systems perform well with essentially zero hallucinations, though \docetl has a marginally higher rate of 0.001 on average.} 
The difference between LOTUS' reported performance and our reimplementation may be attributed to model choices and prompting strategies, as we standardized on gpt-4o-mini without ``few-shot'' examples for consistency across all experiments.

\topic{Cost and Dataset Analysis} \revision{The non-LLM baseline takes 290.65 seconds to run.} Our reimplemented LOTUS pipeline costs \$0.47 and takes 925 seconds to run. The \docetl pipeline costs \$3.65 and takes 463.28 seconds, with an additional optimization cost of \$2.37. 
\techreport{We believe this additional cost is an acceptable tradeoff because
of the between 30-80\% improvement in RP; moreover, these costs
would be insignificant with open-source LLMs. The runtime can be highly variable; as LLMs can have high tail latencies, and LOTUS and \docetl may implement LLM retry logic differently.} 

\techreport{Manual inspection of the results reveals inherent challenges with the task and dataset quality. We found numerous cases where ground truth labels were not actually discussed in the article text, as well as instances where \docetl correctly identified adverse reactions present in the text but missing from the ground truth annotations. This suggests that dataset quality may be the primary factor limiting performance scores across all approaches, rather than limitations of the systems themselves.}

\begin{figure*}
    \centering
    \vspace{-35pt}
    {\includegraphics[width=0.9\linewidth]{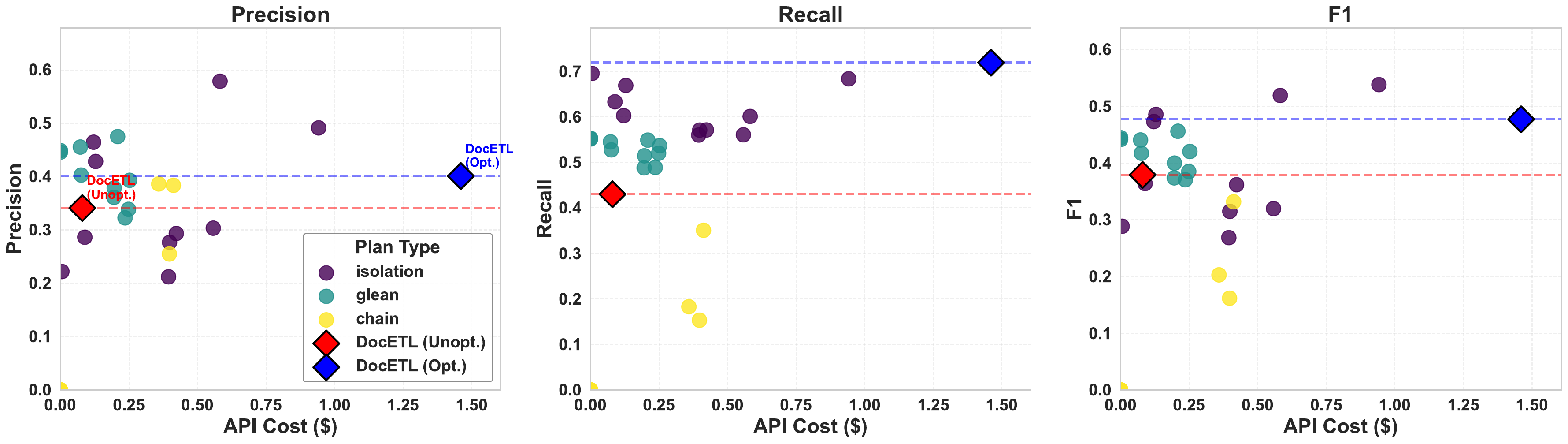}
    }
    \caption{\revision{Cost vs. metrics (precision, recall, and F1) for 30 different LLM-generated implementations of rewrite directives applied to the legal contract analysis task. Each point represents a distinct plan implementation, colored by directive type; isolated projections (\Cref{eq:parallelproj}, chaining projections (\Cref{eq:chainproj}, or gleaning (\Cref{eq:gleaningrewrite}). The \docetl unoptimized baseline and optimized plan from \Cref{sec:evaluation-legal} are shown with dashed lines for reference, though not generated in this experiment. Due to the optimizer's nondeterministic nature, some plans in this experiment achieved higher metrics than the original optimized plan.}}
    \label{fig:legalmultiplegen}
    \vspace{-10pt}
\end{figure*}

\techreport{\subsection{Case Study: Police Misconduct}
\label{sec:evaluation-misconduct}

We conducted an case study on police misconduct identification (\Cref{ex:journalist-task}) using a dataset of 227 documents from various California police departments. This is only a sample of the hundreds of thousands of documents collected by our collaborators at the California Police Records Access Project\footnote{\url{https://cdss.berkeley.edu/news/state-funds-development-first-its-kind-police-misconduct-database}}. This dataset presented several challenges: documents averaged 12,500 tokens, with 2\% exceeding the 128,000 token context window limit. The corpus had an unknown number of cases and several hundred police officers mentioned\footnote{Due to the presence of PII, and the sensitive nature of these documents, we unfortunately cannot open-source our data.}. 

The task was to generate detailed misconduct summaries for each officer who exhibited misconduct, including the officer's name, misconduct types, and a comprehensive summary. We implemented an initial pipeline in \docetl consisting of a {\em map} operation to extract officers who exhibited misconduct from each document, followed by an {\em unnest} operation to flatten this list of officers, and a {\em reduce} operation to summarize misconduct across relevant documents for each officer. For documents exceeding the context limit, we truncated tokens from the middle until they fit within the LLM's context window. Prompts for this pipeline define ``misconduct'' and are written by engineers and journalists employed full-time by the Police Records Access Project.

Running this pipeline as-is led to very incorrect outputs, as police officer names need to undergo entity resolution prior to the reduce operation. In practice, the team runs a domain-specific clustering algorithm, followed by human annotation, to de-duplicate police officer names. As such, our initial pipeline (denoted {\bf Baseline}) therefore also includes a resolve operation before the reduce operation, as per the rewrite directive, \Cref{eq:resolvereduce}. This resolve operation was synthesized by \docetl (i.e., comparison prompt, resolution prompt, and embedding thresholds for blocking).

We evaluated two other pipeline variants, each of which were considered by the optimizer, as well as the final one chosen by the optimizer, all using GPT-4o-mini. It is not obvious which pipeline will be most accurate. The pipelines are as follows:
\begin{enumerate}
    \item {\bf $\docetl_{S}$:} This pipeline applies \Cref{eq:mapreduceproj}---a projection synthesis rewrite---to extract misconduct summaries for identified officers in addition to the officer name before the resolve step. The reduce operation then only summarizes these extracted summaries, as opposed to processing the entire documents.
    \item {\bf $\docetl_{T}$:} This pipeline builds upon $\docetl_{S}$ by extracting both misconduct summaries and types from each document. It then incorporates both the summaries and types in the reduce step, providing more structured information for aggregation.
    \item {\bf  $\docetl_{O}$:} This pipeline, selected by the optimizer, extends $\docetl_{T}$ by chunking documents into 12,840 token segments. It includes metadata extraction and a peripheral context configuration of two previous chunks in full and a summary of earlier content. The map operation is applied to each chunk, followed by a synthesized operation to reduce chunk results per document. Like other pipelines, this is then followed by the officer name resolution step and a final reduce step to aggregate summaries per officer. We will discuss the details of the plan subsequently.
\end{enumerate}

\topic{Results} To evaluate output quality without ground truth data, we came up with three binary criteria: {\em (i)} whether each officer name referred to a real person, {\em (ii)} if the summary included dates and locations of misconduct, and {\em (iii)} whether each identified misconduct instance was extensively described in the summary. To assess the accuracy of our evaluation criteria, we employed GPT-4o-mini as a judge to evaluate each criterion for over 1,500 outputs across the baseline and all variants. To validate the LLM's judgments, we conducted a human evaluation on a subset of the data. For the first two criteria (officer name validity and inclusion of dates/locations), one of the authors manually assessed 100 randomly sampled outputs from both the baseline and DocETL variants. For the third criterion (extensive description of misconduct), due to the detailed and often graphic nature of the summaries, the author evaluated 50 output summaries, a process that required several hours of careful reading. The human evaluation revealed high agreement between the LLM judge and human assessor---96\%, 97\%, and 92\% respectively---suggesting that our LLM-based evaluation method is a reliable proxy for human judgment in this task.

\Cref{tab:misconduct-metrics} illustrates these results. {\bf \em $\docetl_O$ is, on average, $\mathbf{1.34\times}$ more accurate compared to the baseline.} The $\docetl_S$ and $\docetl_{T}$ pipelines performed similarly, with the notable exception of $\docetl_{S}$, which often omitted dates and locations from summaries.

\begin{table}[htbp]
\footnotesize
\centering
\caption{Evaluation Metrics for Police Misconduct Identification Pipelines. Each value represents the fraction of outputs that pass the metric.}
\label{tab:misconduct-metrics}
\begin{tabular}{p{3.5cm}p{0.75cm}p{0.75cm}p{0.75cm}p{0.75cm}}
\toprule
Metric & Baseline & $\docetl_S$ & $\docetl_T$ & $\docetl_O$ \\
\midrule
The officer's name is a specific name, not generic (e.g., not ``Officer 1'') & 0.84 & {\bf 0.93} & 0.89 & 0.87 \\
\addlinespace
The summary contains a date and location & 0.67 & 0.1 & 0.91 &  {\bf 0.92} \\
\addlinespace
Each identified instance of misconduct is described extensively in the summary & 0.42 & 0.78 & 0.76 & {\bf 0.80} \\
\bottomrule
\end{tabular}
\end{table}

Our evaluation underscores the complexity and task-specific nature of assessing LLM-based pipelines. While the outputs of different plans may appear similar at first glance, our analysis reveals some variations in their quality and reliability. The baseline's poor performance highlights the importance of our rewrite rules. $\docetl_S$'s summaries consistently failed to mention locations. $\docetl_T$ and $\docetl_O$ offered the most reliable results, with the latter being particularly suited for longer documents. This variability in plan performance emphasizes the necessity of \docetl's custom validation agents, which demonstrated proficiency in understanding the task-specific nature of evaluation: for instance, the map operation's evaluation prompt focused on the completeness of incident details and correct categorization of misconduct types, while the reduce operation's prompt emphasized accuracy of aggregation and information retention across cases. Without such tailored validation mechanisms, discerning the relative strengths of each plan would be challenging, if not impossible---highlighting the critical role of task-specific optimization and evaluation in LLM-powered document analysis.

\topic{DocETL's Optimized Pipeline} The $\docetl_O$ pipeline
can be expressed using our rewrite directive syntax as follows:
\begin{equation*}
\begin{aligned}
&\text{Map} \to \text{Unnest} \to \text{Reduce} \Rightarrow \\ 
&\begin{aligned}
\text{Map}_M \to \text{Split} \to (\text{Map}_S \parallel \text{Map}_H) \to \text{Gather} \to \text{Map} \to \\
(\text{Map}_v \to \text{Map}_i)^{\leq 1} \to \text{Reduce}_D \to \text{Unnest} \to \text{Resolve} \to \text{Reduce}
\end{aligned}
\end{aligned}
\end{equation*}
where $\{\ttt{officer\_name}\}$ is the reduce key for the final summarization.

This pipeline begins with a map operation to extract metadata ($\text{Map}_M$), followed by document chunking of 12840 tokens each (Split). Each chunk then undergoes \revision{Directive \Cref{eq:summarization}}: $\text{Map}_S$ for summarization and $\text{Map}_H$ for header extraction. The Gather operation collects context for each chunk, including the header lineage for the current chunk, 2 full previous chunks, and summaries of the other previous chunks. The original Map operation is then applied to each rendered chunk, with gleaning applied for refinement. Results from all chunks of a document are combined using $\text{Reduce}_D$. The pipeline then flattens the results (Unnest), resolves officer names (Resolve), and finally summarizes misconduct per officer (Reduce). Overall, this optimized pipeline incorporates several of our rewrite rules, including document chunking (\ref{eq:chunking}), header lineage context and summarization (\ref{eq:summarization}), gleaning for the Map operations (\ref{eq:gleaningrewrite}), and duplicate key resolution (\ref{eq:resolvereduce}).

\topic{Costs} For our sample dataset of 227 documents, the baseline incurred \$2.24, while $\docetl_S$ and $\docetl_T$ each cost \$0.55. $\docetl_O$ was more expensive at \$1.35 due to processing all document chunks, but less expensive than the baseline (due to not needing to include entire documents in the reduce operation). Running the optimizer incurred a cost of approximately \$100 and took about 20 minutes, with the bulk of the expense attributed to validation agents processing lengthy documents. $\docetl_O$ took 364.97 seconds to run, and all other pipelines completed in less than 180 seconds. While the optimization cost of \$100 for a task that takes $< \$3$ may seem high, note that we are merely operating on a sample of the overall dataset; processing the dataset has already cost the team over \$50,000; so this one-time cost of \$100 is amortized across processing hundreds of thousands of documents.  
As part of this process, the optimizer considered and evaluated over 200 pipeline variants. 
As models become more cost-effective (e.g., GPT-4o-mini is over $100\times$ cheaper), optimization costs will decrease significantly, making the investment even more worthwhile in the long run.}

\techreport{\subsection{Case Study: Rewrite Directives in Legal Contract Analysis}
\label{sec:evaluation-legal-casestudy}

One may wonder: if \docetl's agentic optimizer depends on how effectively LLMs transform abstract rewrite directives into concrete plans, how well do they actually perform? We examined this with a case study on our legal contract analysis task from \Cref{sec:evaluation-legal}. We selected three rewrite directive types that don't require physical parameter tuning: projection chaining (synthesizing a chain of dependent map operations, according to Equation \ref{eq:chainproj}), isolated projection (synthesizing independent map operations, per Equation \ref{eq:parallelproj}), and gleaning (Equation \ref{eq:gleaningrewrite}). Using our GPT-4o-powered LLM agent, we generated 10 different instantiations of each directive, yielding 30 distinct plans with varying prompts, structures, and decomposition strategies.

Our study had two objectives: {\em (i)} to assess the quality of LLM-implemented directives and {\em (ii)} to evaluate our LLM-based plan evaluation mechanism. We executed all 30 plans on the 50 contracts, measuring precision, recall, F1 scores, and cost to characterize performance distribution across implementations of the same directive. We then had our optimizer (using GPT-4o-mini as judge) rank these plans per the approach in \Cref{sec:optimizer}, and computed the Kendall's Tau between these rankings and actual performance metrics.

\Cref{fig:legalmultiplegen} illustrates the results. We start with {\em (i)}. Projection synthesis rewrites (both chaining and isolation) showed high variance in both cost and accuracy metrics. Cost variance was expected: more projections synthesized means more LLM calls and proportionally higher costs. More surprising was the substantial accuracy variance, which stemmed from implementation inconsistencies. 20\% of the generated plans (all from chain projection implementations) failed to include document placeholders in prompts (e.g., writing \ttt{extract X clause from the document} instead of \ttt{extract X clause from \{\{ input.document \}\}}), resulting in the LLM receiving no text to analyze. Isolated projections consistently outperformed chaining projections for this task, likely because clause extraction subtasks are mostly independent. Gleaning directive implementations showed more consistent cost profiles, which is expected since they involve exactly three LLM calls per document (initial map, validation, and refinement). Despite implementation variance, many LLM-generated plans outperformed the unoptimized \docetl baseline, with 47\% of plans achieving better precision, 67\% of them achieving better recall, and 40\% of them achieving better F1 scores. When compared against our optimized plan from the main experiment, 30\% still achieved better precision, 10\% achieved better F1 scores, though none improved on recall. Moving on to {\em (ii)}, the Kendall's tau correlation between LLM judge rankings and actual F1 score rankings was 0.642. So, while there is considerable variance in implementation quality, our selection mechanism can identify the more effective plans.

Overall, while LLMs can translate abstract directives into working plans, quality varies considerably, and naive implementations may contain critical errors. This underscores the importance of generating many plans and our LLM-based plan evaluation mechanism and suggests potential improvements through better prompt engineering or selecting models particularly adept at prompt rewriting.}

\techreport{\subsection{User Adoption and Impact}
\label{sec:eval-adoption}
Since releasing \docetl as open source software in October 2024, we have observed increasing adoption across diverse domains and use cases. Users report successfully applying \docetl to complex document processing tasks where other tools struggled---for instance, one user switched from LlamaIndex to \docetl for automatically constructing knowledge graphs from textbooks, citing significantly improved results {\em ``on the first try.''} In another use case, the CEO of a security company who uses \docetl for multiple pipelines (e.g., analyzing logs) said, {\em ``DocETL is simply amazing. It simplifies what would otherwise be a painful document processing pipeline.''} We've seen successful deployments spanning healthcare (medical record analysis), legal (real estate closing documents, regulatory compliance), and scientific research (synthetic biology literature analysis, climate action plan evaluation). \techreport{Users have also applied \docetl to more general enterprise tasks like summarizing customer support tickets, extracting insights from financial reports, and resolving product entities in e-commerce catalogs.} A particularly challenging use case involves forensic psychiatry records dating back to the 1970s, where \docetl effectively processes a mix of handwritten notes, various form formats, and evolving electronic records. While in many cases, \docetl adds multiple synthesized operations to a pipeline, users are able to---and want to---understand \docetl's optimized plans; the operations are simply described by natural language prompts, making them intuitive and transparent.

Since the initial release, based on user feedback, we have extended \docetl with production-critical features including rate limiting and open source model support, optimization of the resolve operator to skip redundant comparisons via transitivity, logging of intermediate outputs and prompts for observability, and a UI playground for rapid prototyping.}

%% file: sections/discussion.tex
\techreport{\section{Discussion and Limitations}
\label{sec:discussion}

We reflect on the differences between DocETL and traditional database systems. Then, we acknowledge limitations of our work.

\topic{Revisiting Traditional Database Paradigms} LLM-powered document processing systems like \docetl differ from traditional database systems across several layers, ranging from operators---both logical and physical---to rewriting and optimization, to user specification and intent. The core difference is that {\em LLMs are inherently non-deterministic and not always accurate.} At the operator level, traditional database systems guarantee correct and consistent outputs, regardless of physical implementation choices. A hash join or nested loop join will produce identical results, differing only in performance characteristics. In contrast, \docetl's output quality and correctness can vary dramatically based on physical operator parameter choices. For example, different fold batch sizes or gather configurations directly impact accuracy rather than just performance. Moreover, even logical operators work differently; unlike traditional DBMSes where atomic attributes (e.g., a blob of text) remain atomic, \docetl must decompose these attributes, because LLM accuracy often decreases as document size increases. The processing order of these decomposed units matters---LLMs cannot always process chunks independently, which motivates specialized operations like our gather operator and folding approach for reduce operations.

Second, rewriting in traditional databases is purely logical and guaranteed to produce identical output, but in \docetl, it's semantic and instantiated by LLMs. In \docetl, operator output quality for rewrites can vary---e.g., if the LLM agent happens to synthesize bad prompts---and rewriting is not just for performance (latency), but also correctness. Moreover, unlike traditional DBMSes where there is only one way of executing a rewrite, here there are infinite. Consequently, the space of possible plans is infinite, even for a single operator, because each plan is instantiated by LLMs that could generate countless variations of prompts and configurations.

Third, optimization in \docetl is different because we lack a good model for accuracy, unlike well-understood cost models in traditional DBMS optimizers. \docetl must empirically evaluate plans using LLMs themselves as judges. LLMs are inherently uncalibrated for accuracy estimation~\cite{liu2024aligning}, so we must score and rank actual sample outputs using pairwise comparisons. Moreover, in \docetl, operator accuracy depends heavily on accuracy of other operators, and during optimization, the best suffix of a plan directly depends on which subplan was chosen for the prefix. This means that unlike traditional optimizers, e.g., Selinger's algorithm~\cite{selinger1979access} or Cascades~\cite{Graefe1995TheCF}, that can optimize subplans and apply dynamic programming or divide and conquer to assemble an optimal plan, \docetl cannot rely on operator independence.

Finally, the specification layer in \docetl is inherently fuzzy. \docetl specifications can be ambiguous because they are in natural language, allowing flexibility in interpretation by LLMs. However, this flexibility comes with limitations---specifications may not capture all corner cases that might arise during processing, requiring robust fault tolerance mechanisms to handle unexpected scenarios when output schemas fail or rate limits are exceeded.

\topic{Limitations} While \docetl demonstrates accuracy improvements in real-world applications, there are some limitations. First, using LLMs for both directive implementation and validation risks shared biases, although our validation agent's different perspective often identifies issues despite sharing the underlying model. Recent work from the ML community has demonstrated that LLMs can often be more effective as verifiers than generators, especially if prompted to focus on specific criteria, as our validation agents do~~\cite{zheng2023judging, zhujudgelm, zhang24generative}. Nevertheless, our optimizer allows users to specify different models for the rewrite and validation agents, and we're exploring several other mitigation strategies: diversifying prompts; incorporating human expertise in the optimizer through interactive refinement of pipelines; and exploring hybrid validation approaches that {\em combine} LLM assessment with traditional techniques (e.g., allowing a user to provide labeled ground truth to verify accuracy via exact string matching). These ideas are guiding our development of a prototype interface for \docetl that enables users to incorporate their expertise both during optimization and by directly editing or correcting outputs, providing a human-in-the-loop approach to complement automated complex document processing.}

%% file: sections/related.tex
\section{Related Work}\label{sec:related}

LLM-powered data processing frameworks have recently gained significant attention in the database community. LOTUS~\cite{patel2024lotus} extends Pandas with semantic operators, while Palimpzest~\cite{liu2024declarative} provides a declarative framework focusing on map-like operations. Aryn~\cite{anderson2024designllmpoweredunstructuredanalytics} offers a Spark-like API with PDF extraction capabilities and human-in-the-loop processing.  \revision{Unlike \docetl, these systems primarily make simplifying assumptions about task complexity, typically focusing on extraction tasks or queries that capable LLMs can handle without decomposition. They employ various cost-based optimizations, including classical techniques like predicate pushdown~\cite{hellerstein2005anatomy} and ML-specific approaches like model cascades~\cite{wang2017idk}. However, when applied to complex document processing tasks, even state-of-the-art models fall short.} \docetl addresses this limitation through agent-driven optimization, exploring decomposition to improve accuracy. Moreover, \docetl is, uniquely, the only system to support documents with lengths that exceed LLM context windows \revision{and introduces new operators (e.g., gather, split) for this, as well as for entity resolution as a first-class citizen}.

Other LLM-powered data processing systems focus on different settings\revision{---typically making strong assumptions about the structure of the documents and predictability of format}. ZenDB~\cite{lin2024towards} optimizes SQL queries for templatized documents, while \docetl handles arbitrary documents. EVAPORATE~\cite{arora2023language} specializes in table extraction through code synthesis (only where applicable, in semi-structured settings), which could complement \docetl. Regarding LLM agents: Caesura~\cite{urban2024demonstrating} uses LLMs to translate natural language to SQL pipelines but leaves optimization for future work; CleanAgent~\cite{qi2024cleanagent} uses agents to standardize and clean data (and also does not consider optimization). Other systems propose specialized pipelines for specific tasks: for instance, \citet{edge2024local} use a fixed map-reduce pipeline with predefined prompts for knowledge graph querying---whereas \docetl enables flexible pipeline construction and optimization for any document processing task. \revision{A common limitation across these systems is inadequate context management, particularly for documents exceeding context windows or tasks requiring cross-document reasoning.}  While prompt optimization~\cite{wen2024hard,khattab2024dspy} could complement \docetl, it falls short on complex document tasks, even with human guidance~\cite{white2023prompt}\techreport{, particularly when data and tasks exceed single LLM call capabilities}. Moreover, LLMs have been leveraged for a variety of data tasks beyond document processing, such as join discovery~\cite{dong2022deepjoin,kayali2024chorus}, database tuning~\cite{trummer2022db}, ML pipelines~\cite{shankar2024building}, natural language to SQL~\cite{pourreza2024chase}, semantic table understanding~\cite{fang2024large,cong2023observatory}. and others~\cite{fernandez2023large}, but not for complex document processing.

Finally, declarative frameworks for intelligent data processing have a rich history in database research through crowdsourcing systems like CrowdDB, Deco, CDB, and Qurk~\cite{franklin2011crowddb, parameswaran2012deco, li2018cdb, marcus2011crowdsourced}. While these systems use human rather than machine intelligence, they demonstrate declarative interfaces' power for complex tasks. \docetl extends this tradition to address the unique challenges of LLM-powered processing~\cite{parameswaran2023revisiting} through its flexible interface and agent-driven optimization.

%% file: sections/conclusion.tex
\section{Conclusion}
\label{sec:conclusion}

We introduced \docetl, a declarative system that optimizes complex document processing tasks using LLMs. We introduced several novel rewrite directives, an agent-based framework for plan rewriting and evaluation, and an opportunistic optimization strategy. Our evaluation across four unstructured document analysis tasks demonstrated that \docetl can find plans with outputs 21-80\% more accurate than baselines. \revision{\docetl is a first step toward an agentic optimizer for LLM-powered data processing. While exploring the large space of possible plan decompositions is hard, our approach shows that automated optimization is both feasible and beneficial. Future work will focus on reducing cost by considering cheaper models for simpler sub-tasks, and incorporating human feedback to refine plans. With growing real-world adoption, \docetl provides a foundation for future research and applications in complex document processing.}

%% file: sections/appendix.tex
\section{Gather Operator Specifications}
\label{app:gather}

\subsection{Gather Configuration}

The gather operation's configuration includes:

\begin{itemize}
    \item The group ID key (document ID)
    \item The order key (chunk sequence within a group)
    \item The content key (field containing chunk content)
    \item The peripheral chunk configuration
\end{itemize}

The peripheral chunk configuration specifies ``previous'' and ``next'' sections, each potentially containing ``head'', ``middle'', and ``tail'' subsections, determining which surrounding chunks to include and how many. Each subsection must specify a \ttt{content\_key} denoting the field to use as the content of the chunk.

\subsection{Header Lineage Preservation}

\begin{figure*}
\centering
\caption{Example of Document Header Handling in a Gather Operation for Legal Contracts~\cite{hendrycks2021cuad}. The example document has 74 pages. Headers are extracted from chunks via map operations. When rendering a chunk (e.g., Chunk 20), the operation includes the most recent headers of all levels (1, 2, etc.) above the first header in the current chunk, so the LLM has hierarchical context when processing the chunk.}
\label{fig:headergather}
\includegraphics[width=0.9\linewidth]{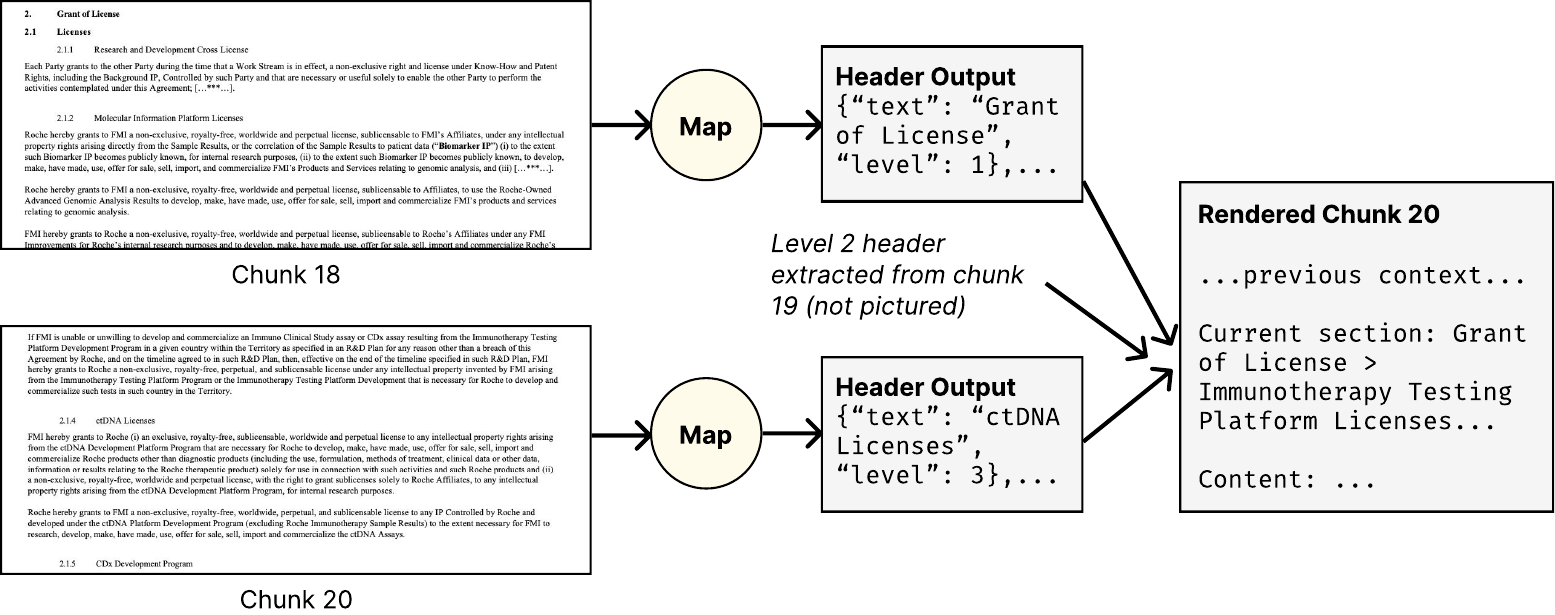}
\end{figure*}

A unique feature of the gather operation is its ability to maintain document structure through headers. This is particularly useful for documents with complex structures where processing a chunk with a certain level header requires knowledge of headers in the levels above, which may be in other chunks.

When a \ttt{doc\_header\_key} is specified in the configuration, the gather operation:

\begin{enumerate}
    \item Examines the \ttt{doc\_header\_key} field for every chunk preceding the one being rendered.
    \item Reconstructs the relevant header structure by identifying the level of the first header in the current chunk and including all most recent headers from higher levels found in previous chunks.
    \item Arranges these headers in their proper order.
\end{enumerate}

This process ensures that each rendered chunk includes a complete ``path'' of headers leading to its content, preserving the document's overall structure and context even when split across multiple chunks.

\Cref{fig:headergather} demonstrates header handling in a gather operation for a 74-page legal contract. Headers are extracted from chunks via map operations. When rendering a chunk (e.g., Chunk 20), the operation includes the most recent headers of all levels (1, 2, etc.) above the first header in the current chunk, providing hierarchical context for LLM processing.

\newpage
\section{Rewrite Directive Instantiation}
\label{app:example-rewrite}

This appendix provides a detailed walkthrough of how an LLM agent instantiates a rewrite directive in \docetl, using the legal contract analysis task in \Cref{sec:evaluation-legal}.

\subsection{Initial Task and Baseline Operation}

Our baseline approach used a single \texttt{map} operation with a prompt listing all 41 clause types to extract from legal contracts:

\begin{lstlisting}[language=text,style=plaintextstyle]
Given the following contract document:

{{ input.document }}

Extract the text spans (if they exist) for each of the following categories. If a category is not present or cannot be determined, return an empty string. If there are multiple text spans for a category, return them as a comma-separated list of text spans.

1. Document Name: The name of the contract
2. Parties: The two or more parties who signed the contract
3. Agreement Date: The date of the contract
4. Effective Date: The date when the contract is effective
5. Expiration Date: On what date will the contract's initial term expire?
6. Renewal Term: What is the renewal term after the initial term expires? This includes automatic extensions and unilateral extensions with prior notice.
7. Notice to Terminate Renewal: What is the notice period required to terminate renewal?
8. Governing Law: Which state/country's law governs the interpretation of the contract?
9. Most Favored Nation: Is there a clause that if a third party gets better terms on the licensing or sale of technology/goods/services described in the contract, the buyer of such technology/goods/services under the contract shall be entitled to those better terms?
10. Non-Compete: Is there a restriction on the ability of a party to compete with the counterparty or operate in a certain geography or business or technology sector?
11. Exclusivity: Is there an exclusive dealing commitment with the counterparty? This includes a commitment to procure all "requirements" from one party of certain technology, goods, or services or a prohibition on licensing or selling technology, goods or services to third parties, or a prohibition on collaborating or working with other parties), whether during the contract or after the contract ends (or both).
12. No-Solicit of Customers: Is a party restricted from contracting or soliciting customers or partners of the counterparty, whether during the contract or after the contract ends (or both)?
13. Competitive Restriction Exception: This category includes the exceptions or carveouts to Non-Compete, Exclusivity and No-Solicit of Customers above.
14. No-Solicit of Employees: Is there a restriction on a party's soliciting or hiring employees and/or contractors from the counterparty, whether during the contract or after the contract ends (or both)?
15. Non-Disparagement: Is there a requirement on a party not to disparage the counterparty?
16. Termination for Convenience: Can a party terminate this contract without cause (solely by giving a notice and allowing a waiting period to expire)?
17. Right of First Refusal, Offer or Negotiation: Is there a clause granting one party a right of first refusal, right of first offer or right of first negotiation to purchase, license, market, or distribute equity interest, technology, assets, products or services?
18. Change of Control: Does one party have the right to terminate or is consent or notice required of the counterparty if such party undergoes a change of control, such as a merger, stock sale, transfer of all or substantially all of its assets or business, or assignment by operation of law?
19. Anti-Assignment: Is consent or notice required of a party if the contract is assigned to a third party?
20. Revenue/Profit Sharing: Is one party required to share revenue or profit with the counterparty for any technology, goods, or services?
21. Price Restriction: Is there a restriction on the ability of a party to raise or reduce prices of technology, goods, or services provided?
22. Minimum Commitment: Is there a minimum order size or minimum amount or units per-time period that one party must buy from the counterparty under the contract?
23. Volume Restriction: Is there a fee increase or consent requirement, etc. if one party's use of the product/services exceeds certain threshold?
24. IP Ownership Assignment: Does intellectual property created by one party become the property of the counterparty, either per the terms of the contract or upon the occurrence of certain events?
25. Joint IP Ownership: Is there any clause providing for joint or shared ownership of intellectual property between the parties to the contract?
26. License Grant: Does the contract contain a license granted by one party to its counterparty?
27. Non-Transferable License: Does the contract limit the ability of a party to transfer the license being granted to a third party?
28. Affiliate IP License-Licensor: Does the contract contain a license grant by affiliates of the licensor or that includes intellectual property of affiliates of the licensor?
29. Affiliate IP License-Licensee: Does the contract contain a license grant to a licensee (incl. sublicensor) and the affiliates of such licensee/sublicensor?
30. Unlimited/All-You-Can-Eat License: Is there a clause granting one party an "enterprise," "all you can eat" or unlimited usage license?
31. Irrevocable or Perpetual License: Does the contract contain a license grant that is irrevocable or perpetual?
32. Source Code Escrow: Is one party required to deposit its source code into escrow with a third party, which can be released to the counterparty upon the occurrence of certain events (bankruptcy, insolvency, etc.)?
33. Post-Termination Services: Is a party subject to obligations after the termination or expiration of a contract, including any post-termination transition, payment, transfer of IP, wind-down, last-buy, or similar commitments?
34. Audit Rights: Does a party have the right to audit the books, records, or physical locations of the counterparty to ensure compliance with the contract?
35. Uncapped Liability: Is a party's liability uncapped upon the breach of its obligation in the contract? This also includes uncap liability for a particular type of breach such as IP infringement or breach of confidentiality obligation.
36. Cap on Liability: Does the contract include a cap on liability upon the breach of a party's obligation? This includes time limitation for the counterparty to bring claims or maximum amount for recovery.
37. Liquidated Damages: Does the contract contain a clause that would award either party liquidated damages for breach or a fee upon the termination of a contract (termination fee)?
38. Warranty Duration: What is the duration of any warranty against defects or errors in technology, products, or services provided under the contract?
39. Insurance: Is there a requirement for insurance that must be maintained by one party for the benefit of the counterparty?
40. Covenant Not to Sue: Is a party restricted from contesting the validity of the counterparty's ownership of intellectual property or otherwise bringing a claim against the counterparty for matters unrelated to the contract?
41. Third Party Beneficiary: Is there a non-contracting party who is a beneficiary to some or all of the clauses in the contract and therefore can enforce its rights against a contracting party?
\end{lstlisting}

The output schema specified a list of clause-text pairs:

\begin{lstlisting}[language=text,style=plaintextstyle]
output:
  schema:
    clauses: 'list[{clause_type: string, text_span: string}]'
\end{lstlisting}

\subsection{Optimization Process}

\subsubsection{Determine If Optimization Is Needed}

The \docetl optimizer first evaluates whether the operation needed improvement. The system uses an LLM agent to synthesize a validator prompt. This agent is provided the operation prompt and sample outputs, and instructed to devise binary validation criteria around precision and recall. The following validator prompt was generated:

\begin{lstlisting}[language=text,style=plaintextstyle]
1. Did the operation successfully extract all relevant text spans for each 
   category listed in the task prompt, ensuring no instances were missed, 
   especially for categories with multiple possible text spans?      
2. Is the extracted information presented in the correct format, with 
   text spans properly separated by commas where applicable, and are 
   there any inconsistencies or errors in the extracted data? 
\end{lstlisting}

When applied to sample outputs from the original user-written operation, the LLM validator replies:

\begin{lstlisting}[language=text,style=plaintextstyle]
"Needs improvement. Row 1 output is mostly accurate, but it misses several important categories such as 'Renewal Term', 'Notice to Terminate Renewal', 'Governing Law', 'Most Favored Nation', 'Exclusivity', 'No-Solicit of Customers',... The extraction process should be improved to ensure that all categories listed in the task prompt are addressed, even if the information is not present in the document. For categories that do not have applicable text spans, the output should explicitly state that the category is not applicable or that the information is missing, rather than omitting it entirely."
\end{lstlisting}

\subsubsection{ Select and Apply Rewrite Directive}

Based on this assessment, \docetl programmatically matches the case to applicable rewrite directives. For this example, we will focus on the implementation of \Cref{eq:parallelproj}, which decomposes a complex map operation into independent subtasks. The optimizer then invokes the LLM agent with: the original prompt and output schema, sample contract documents, and instructions to decompose the task. As of March 2025,\docetl uses the following prompt to the LLM agent:

\begin{lstlisting}[language=text,style=plaintextstyle]
Original task prompt:
{original_prompt}

Output schema the operation will produce:
{
  "clauses": "list[{clause_type: string, text_span: string}]"
}

Input data sample:
{sample_contract_text}

Decompose the original task into parallel subtasks, where each subtask 
produces one or more keys of the output schema. Assume that the subtasks 
can be executed independently. You cannot rely on the output of one subtask 
to complete another subtask. Make sure you include the same input variables 
as in the original task prompt. Each prompt should be a Jinja2 template...

[Additional instructions about format...]
\end{lstlisting}

\subsubsection{LLM Agent Generates Decomposed Operations}

The LLM agent responded with a detailed decomposition plan:

\begin{lstlisting}[language=text,style=plaintextstyle]
{
  "subtasks": [
    {
      "name": "extract_basic_contract_information",
      "prompt": "From the following contract document: {{ input.document }}\nExtract the following basic information:\n- Document Name: The name of the contract\n- Parties: The two or more parties who signed the contract\n- Agreement Date: The date of the contract\n- Effective Date: The date when the contract is effective\n- Expiration Date: On what date will the contract's initial term expire?",
      "output_keys": [
        "document_name",
        "parties",
        "agreement_date",
        "effective_date",
        "expiration_date"
      ]
    },
    {
      "name": "extract_renewal_and_termination_clauses",
      "prompt": "From the following contract document: {{ input.document }}\nExtract the following clauses related to renewal and termination:\n- Renewal Term: What is the renewal term after the initial term expires?\n- Notice to Terminate Renewal: What is the notice period required to terminate renewal?\n- Termination for Convenience: Can a party terminate this contract without cause?",
      "output_keys": [
        "renewal_term",
        "notice_to_terminate_renewal",
        "termination_for_convenience"
      ]
    },
    {
      "name": "extract_legal_governance_clauses",
      "prompt": "From the following contract document: {{ input.document }}\nExtract the following legal governance clauses:\n- Governing Law: Which state/country's law governs the interpretation of the contract?\n- Anti-Assignment: Is consent or notice required of a party if the contract is assigned to a third party?\n- Change of Control: Does one party have the right to terminate or is consent or notice required if such party undergoes a change of control?",
      "output_keys": [
        "governing_law",
        "anti_assignment",
        "change_of_control"
      ]
    },
    ....further clauses omitted due to brevity
  ]
}
\end{lstlisting}

We similarly prompt the LLM to synthesize an additional operation to convert the flat schema output from the \ttt{parallel\_map} into the original output schema format expected by the user, following \Cref{eq:parallelproj}. For the sake of this walk-through, we will assume that each newly synthesized operation is able to accurately perform its subtask, but in practice, \docetl will recursively apply rewrites to each new subtask.

\subsubsection{Plan Evaluation and Selection}

After generating the aforementioned plan and several alternatives (with variations in groupings and decomposition strategies), \docetl executes all the candidate plans on a sample, and invokes the validation agent to rate and compare outputs of plans. \docetl picks the top-ranked plan to replace the original operation with. Overall, the LLM agent effectively turns a high-level rewrite directive into a concrete implementation by, understanding the original task and data, identifying semantically meaningful ways to decompose it, generating appropriate prompts for each subtask, and creating the necessary schema transformations.